\documentclass[useAMS,usenatbib]{mnras}
\usepackage{aas_macros}
\usepackage{graphicx}
\usepackage[usenames]{color}
\usepackage{times}
\def\gtsim{\lower.5ex\hbox{$\; \buildrel > \over \sim \;$}}
\def\lesssim{\lower.5ex\hbox{$\; \buildrel <\over \sim \;$}}

\title[Green valley galaxy transformation]{Galaxy and Mass Assembly (GAMA): Morphological transformation of  galaxies across the green valley}

\author[ M.N. Bremer et al.]
{M.N. Bremer$^1$, S. Phillipps$^1$,  L.S. Kelvin$^2$,  R. De Propris$^3$, 
Rebecca Kennedy$^4$, \and
Amanda J. Moffett$^5$, S. Bamford$^4$, L.J.M. Davies$^5$,  S. P. Driver$^{5,6}$,  B. H\"{a}u{\ss}ler$^7$, \and  B. Holwerda $^8$, A. Hopkins$^9$, P.A. James$^2$,  J. Liske$^{10}$, S. Percival$^2$, E.N. Taylor$^{11}$\\
$^1$HH Wills Physics Laboratory, University of Bristol, Tyndall Avenue, Bristol, BS8 1TL, UK\\
$^2$ Astrophysics Research Institute, Liverpool John Moores University, 146 Brownlow Hill, Liverpool L3 5RF, UK\\
$^3$ Finnish Centre for Astronomy with ESO, University of Turku, Vaisalantie 20, Piikkio, Finland\\
$^4$ School of Physics and Astronomy, The University of Nottingham, University Park, Nottingham NG7 2RD, UK\\
$^5$ International Centre for Radio Astronomy Research, University of Western Australia, M468, 35 Stirling Highway, Crawley, WA 6009, Australia\\
$^6$ (SUPA) School of Physics \& Astronomy, University of St. Andrews, North Haugh, St. Andrews, Fife, KY16 9SS, UK.\\
$^7$ESO - European Southern Observatory, Alonso de Cordova 3107, Vitacura, Santiago, Chile\\
$^8$ Department of Physics and Astronomy, University of Louisville, 102 Natural Science Building, Louisville, KY 40292, USA\\
$^9$ Australian Astronomical Observatory, PO Box 915 North Ryde, NSW 1670, Australia\\
$^{10}$ Hamburger Sternwarte, Universit\"{a}t Hamburg, Gojenbergsweg 112, 21029 Hamburg, Germany\\
$^{11}$ Centre for Astrophysics and Supercomputing, Swinburne University of Technology, Hawthorn, Vic 3122, Australia
}

\begin{document}

\date{Accepted 2017 December 18. Received 2017 December 14 ; in original form  2017 July 26}

\pagerange{\pageref{firstpage}--\pageref{lastpage}} \pubyear{2017}

\maketitle

\label{firstpage}

\begin{abstract}

We explore constraints on the joint photometric and morphological evolution of typical low redshift galaxies as they move from the blue cloud through the green valley and onto the red sequence. We select GAMA survey galaxies with $10.25<{\rm log}(M_*/M_\odot)<10.75$ and $z<0.2$ classified according to their intrinsic   $u^*-r^*$ colour. From single component S\'{e}rsic fits,  we find that the stellar mass-sensitive $K-$band profiles of red and green galaxy populations are very similar, while $g-$band  profiles indicate  more disk-like morphologies for the green galaxies: apparent (optical) morphological differences arise primarily from  radial mass-to-light ratio variations. Two-component fits show that most green galaxies have significant bulge and disk components and that  the blue to red evolution  is  driven by colour change in the disk. Together, these strongly suggest that galaxies evolve from blue to red through secular disk fading and that a strong bulge is present prior to any decline in star formation. The relative abundance of the green population implies a typical timescale for traversing the green valley $\sim 1-2$~Gyr and is independent of environment, unlike that of the red and blue populations. While environment likely plays a r\^ole in triggering the passage across the green valley, it appears to have little effect on  time taken. These results are consistent with a green valley population dominated by (early type) disk galaxies that  are insufficiently supplied with gas to maintain previous levels of disk  star formation, eventually attaining passive  colours. No single event is needed quench their star formation.

\end{abstract}


\section{Introduction}

Once the dichotomy of `red' and `blue' galaxy populations was firmly established in its modern context  \cite[][
{\it et. seq.}]{Strateva2001,Baldry2004,Driver2006}, after earlier work by \cite{Tully1982}, it took little time to establish the existence of a rarer population of galaxies falling between the two populations in colour magnitude diagrams - the so-called `green valley' population \citep{Martin2007, Wyder2007}. These galaxies have subsequently been studied as likely examples of systems in transition from actively star forming to passively evolving  (e.g. \citealt{Fang2013} and the review by \citealt{Salim2014}), although it has also been argued that this population is the overlap of two broad `red' and `blue' distributions \citep{Taylor2015} or that it can include previously passive objects where star formation has been rejuvenated \citep[e.g.][]{Thilker2010, Mapelli2015}. The observation by \cite{Bell2007} that at $z<1$ stars form predominantly in blue galaxies, but stellar mass accumulates in red galaxies (see also, e.g., \citealt{Eales2015}) demonstrates the importance of studies of  galaxies transitioning from blue to red populations in our understanding of the last $\sim 7$~Gyr of galaxy evolution  \citep[][and many others]{PengY2010, Lilly2013}.

Recent papers have pointed out that there is a continuum of specific star formation rates (SSFR) in galaxies that do not necessarily show a clear ``valley" between the passive and strongly star forming populations \citep{Oemler2017, Eales2017a, Terrazas2017}. At the extremes of the SSFR distribution the range of colours is limited by the stellar populations present in such galaxies, so galaxies tend to ``pile up" at the ends of the ranges, generating a valley in between when viewed in colour space. Nevertheless galaxies with intermediate SSFRs will generally have stellar populations with intermediate colours and clearly objects transitioning from high to low SSFRs will pass through this region of colour space.

A significant  issue with interpreting green  galaxies as being in transition is the apparently clear difference in morphology between the blue and red populations. Classically, the blue galaxies are identified with star forming disk-dominated systems and the red population with spheroids dominated by passively evolving stellar populations. For example \cite{Schawinski2014} demonstrates that the low redshift green valley  contains objects with both types of morphology when classified in UV and optical bands, with the mix of apparent morphologies switching from disk-dominated to spheroids as the green valley is traversed from blue to red in UV-optical colour. If treated as two separate populations within the valley, the timescale and mechanism for truncating or reducing star formation would have to be different in the two morphological populations.

However, if the entire green valley population (including the significant fraction of galaxies with indeterminate morphology) is considered as a single one which evolves in both apparent morphology and colour (e.g. via disk fading), a different conclusion may result \citep[e.g., see][]{Casado2015}.  Due to the variation in the morphological mix of samples with  mass, it is clearly important to  take account of the mass ranges used in such work.  \cite{Fang2013} demonstrated that the central stellar mass density distributions  of green valley galaxies  and red passively evolving systems of the same total stellar mass occupy the same region of parameter space.  A typical member of the blue population at the same stellar mass had significantly lower central densities, with only a small subset having central surface densities approaching those of the red and green populations.  Their interpretation was that the formation of a dense central spheroid was a necessary step that precedes termination or significant reduction of star formation, implying that morphological transformation is already underway before quenching, assuming that the green galaxies are transitioning redwards \citep[see also, e.g., ][]{Bluck2014}. However, it is evident that this result is also consistent with rejuvenation of spheroid dominated `red' objects.

Here we take the opportunity afforded by the availability of the rich multi-waveband data set for the highly spectroscopically complete GAMA survey of low redshift galaxies   to explore this potential transitional population. All basic galaxy parameters used in this work were previously derived and verified in earlier GAMA-based studies and were taken from the GAMA database \citep{Driver2011, Liske2015}. In particular we exploit the availability  of simple structural measurements made in  near-IR wavebands in order to explore how much of the perceived difference between  blue, green and red populations is due to fundamentally different radial mass distributions as opposed to variations in stellar populations and/or star formation, which may dominate at shorter wavelengths. 

In this work, we specifically exploit the relative stability of the $K-$band mass to light ratio against variation in star formation history \citep{Bell2003}. Given that the $K-$band light is predominantly emitted by red dwarfs and post main-sequence red giants, it more closely traces the underlying stellar mass distributions compared to the  optical and UV morphology (though see Section 3 for caveats on this). A star forming disk may dominate in the optical and UV (leading to a late-type classification for a galaxy) whereas it may only contain a small fraction of the stellar mass. In an extreme case, an object may appear disk-dominated in the optical or UV and as a spheroid in the near-IR, which more accurately reflects the distribution of mass. Similar arguments have been made by \cite{Fang2013} for relating central surface brightness in the $i-$band to the surface density of evolved bulge components in low redshift galaxies.

\begin{figure}
\includegraphics[width=\linewidth]{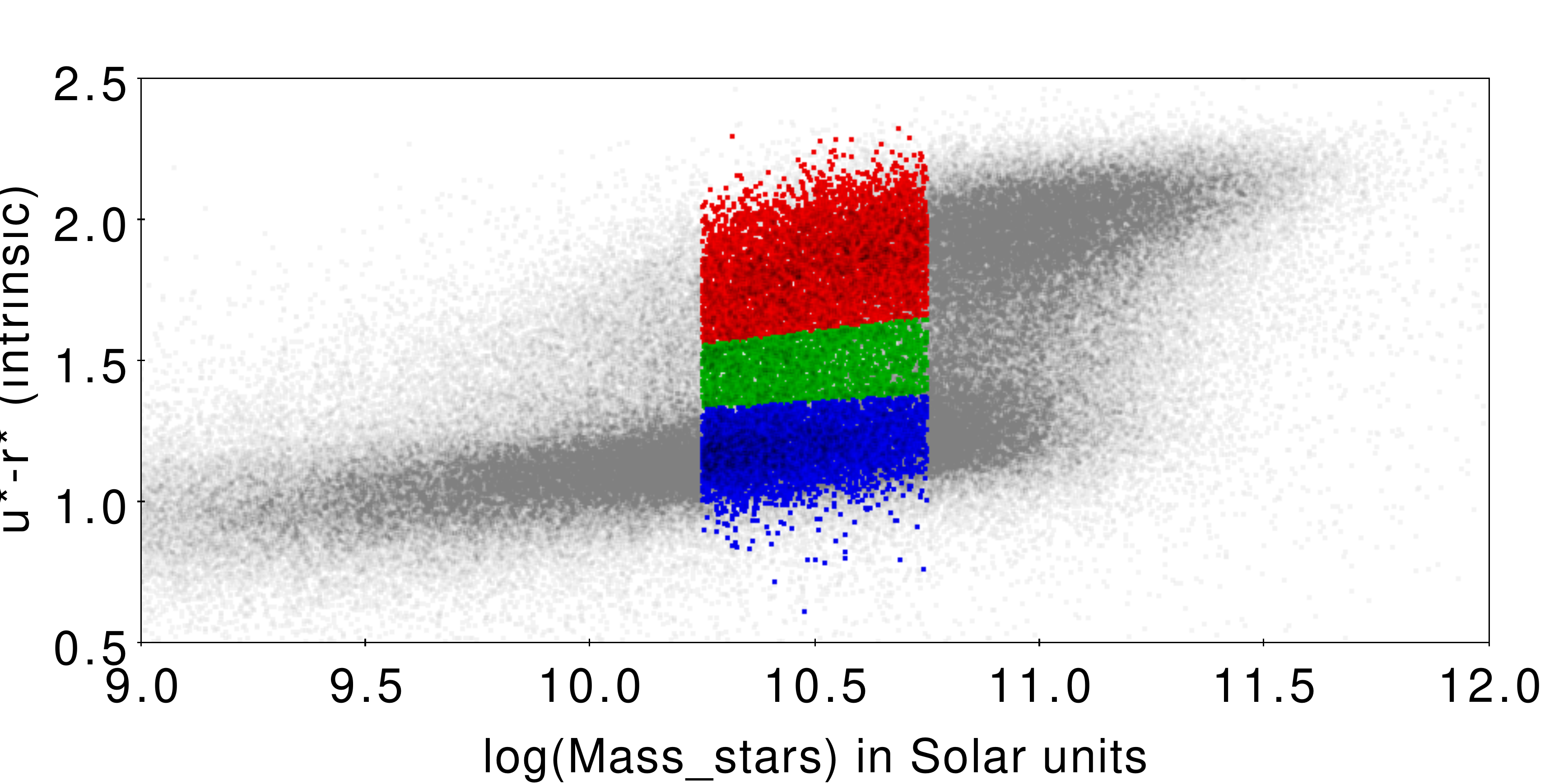}

\caption{The stellar mass vs. intrinsic stellar population colour plane for all $z<0.2$ GAMA galaxies. Overplotted in the appropriate colour are the data for the subsets defined as red, green and blue in the text. These objects have $0.1<z<0.2$, stellar masses of $10.25 <{\rm log} (M_*/M_\odot) <10.75$ and projected axial ratios of $b/a>0.5$.}
\label{CMDa}
\end{figure}

This investigation uses  a stellar mass limited ($10.25<$ log$(M_*/M_\odot)<10.75$, a range that well samples the full range of galaxy colours) and volume complete ($0.1<z<0.2$) sample of just under 10000 galaxies taken from the three equatorial fields of the GAMA-II survey, split into three colour-selected subsamples (red, green and blue) based on the intrinsic (extinction-corrected) optical  colours of their stellar populations. For this sample, rather than attempt to  classify the apparent optical morphology of each (relatively small) image by eye, we prefer to use the available single-component S\'ersic fits to each galaxy in order to quantify the structure of each source, irrespective of apparent optical morphology \citep[cf.][]{Fang2013}. This allows us to probe to higher redshift than possible by eye, thus  providing a large statistical sample and  an objective and, crucially, continuous classification of the sample, avoiding  issues with ``indeterminate" or ``intermediate" objects.  The latter can form a significant fraction or even the majority  of a sample classified by eye at these redshifts using ground-based data of this quality \citep[e.g., as in the Galaxy Zoo based analysis of][]{Schawinski2014}. 

We then investigate measures of morphology for some of the objects within the different colour subsamples in more detail. In particular we use the two-component (bulge and disk) decomposition of objects drawn from one of the equatorial fields by  \cite{Kennedy2016} to explore how the overall colour is affected by those of individual morphological components. We also briefly consider the Hubble type classification by  \cite{Moffett2016} of $z<0.06$ GAMA galaxies in the three equatorial fields, split into red, green and blue subsamples, in order to look at the distribution of Hubble types as a function of colour and compare it to the work on the $0.1<z<0.2$ sample.

All magnitudes used in this work are in the AB system. Where relevant we use H$_0$=70 kms$^{-1}$Mpc$^{-1}$, $\Omega_m=0.3$ and $\Omega_\Lambda = 0.7$ as in \cite{Taylor2011}, from where we take the GAMA masses and luminosities.

\section{Sample selection}
\label{samp}
The GAMA survey provides a multi-wavelength database \citep{Driver2011, Driver2016} based on a highly complete galaxy redshift survey \citep{Baldry2010, Hopkins2013, Liske2015} covering 280 deg$^2$ to a main survey limit of $r_{AB }< 19.8$ mag in three equatorial (G09, G12 and G15) and two southern (G02 and G23) regions. The spectroscopic survey was undertaken using the AAOmega fibre-fed spectrograph \citep{Sharp2006, Saunders2004} in conjunction with the Two-degree Field (2dF, \citealt{Lewis2002}) positioner on the Anglo-Australian Telescope. It obtained redshifts for $\sim 250,000$ targets covering $0<z \lesssim 0.5$ with a median redshift of $z\sim 0.2$ and with highly uniform spatial completeness on the sky \citep{Robotham2010, Driver2011}. Full details of the GAMA survey can be found in \cite{Driver2011} and \cite{Liske2015}. In this work we utilise the first five years of data obtained and frozen for internal team use, referred to as GAMA II. 

Within the multi-wavelength database, each galaxy is characterised by a wide range of  observed and derived parameters, both photometric and spectroscopic. These include stellar masses and intrinsic extinction-corrected  rest-frame colours for the stellar population, derived from multi-wavelength SED fitting \citep{Taylor2011}, star formation rates derived from spectral line analysis \citep{Hopkins2013, Davies2016}, single component two-dimensional S\'ersic fits across all optical and near-IR wavebands \citep{Kelvin2012} and measures of environment \citep{Robotham2011, Brough2013}.  For our main analysis (Section 3) we use a complete set of data drawn from the three equatorial fields (G09, G12, G15). For the later parts of our study we also use the previously-determined bulge-disk decompositions from \cite{Kennedy2016} and morphological classifications from \cite{Moffett2016}, which are available only for limited subsets of GAMA galaxies (discussed in Section 4).

\begin{figure}

\includegraphics[width=\linewidth]{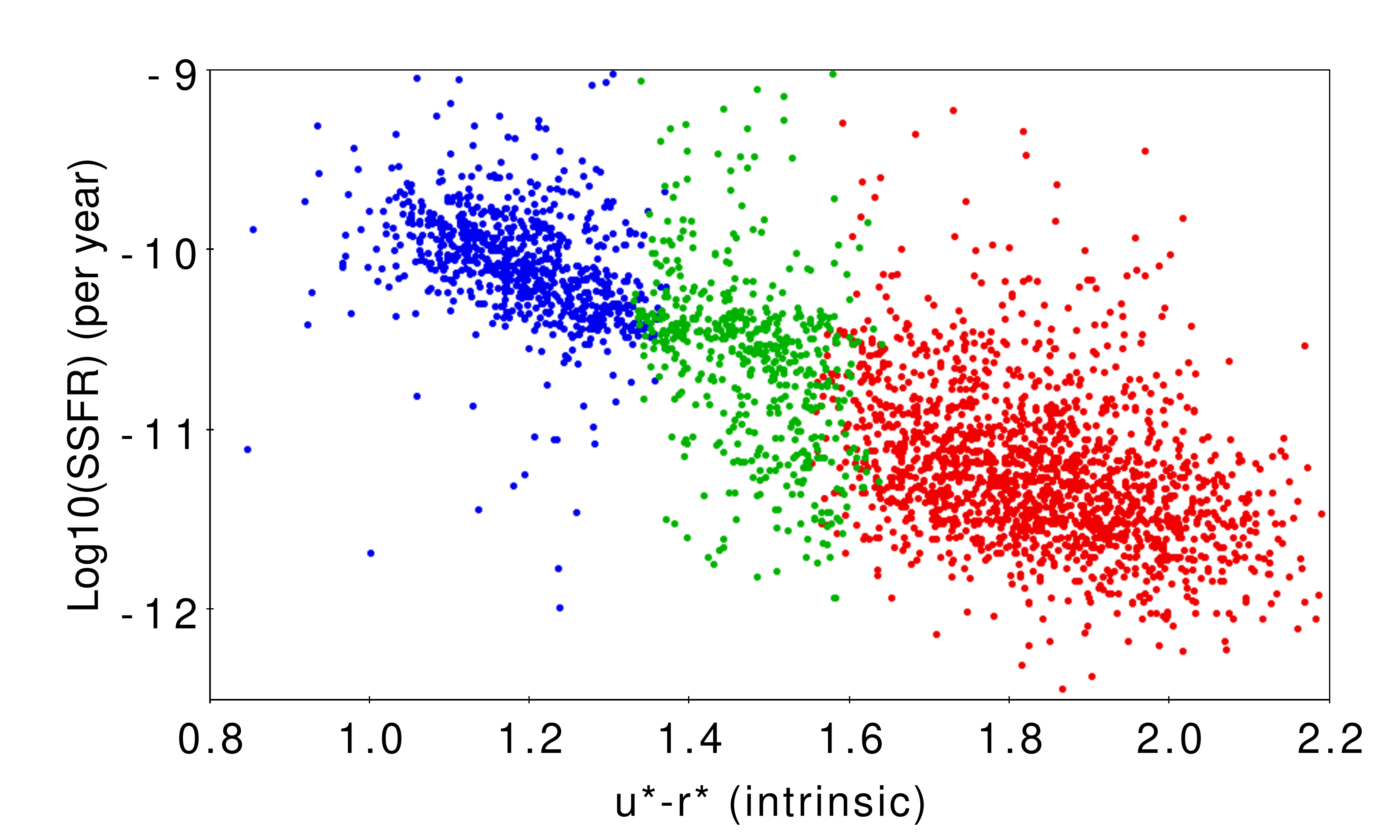}

\caption{ Specific star formation rate  averaged over the last Gyr as a function of intrinsic stellar population colour for the subsets of red, green and blue galaxies defined in the text.   Star formation rates were taken from the GAMA database and determined using the MAGPHYS code \citep{Cunha2008}  on the 21-band photometry of each object (GALEX to WISE) described in  \citet{Wright2016}.}

\label{SSFR}

\end{figure}

For the purposes of the present paper, we  need to select galaxies from a  region  (or regions) of parameter space unbiassed with respect to  the intrinsic colour of a galaxy's stellar population  and with a sufficiently large sample to clearly demonstrate any trend with that colour.  

For our primary sample, used throughout Section 3, we select all galaxies within the three equatorial regions of GAMA with $0.1<z<0.2$ (giving a $\sim 1$~Gyr range in look-back times, minimising any evolutionary effects) and  with stellar masses $10.25 <{\rm log} (M_*/M_\odot) <10.75$.\footnote{From v18 of the GAMA stellar masses catalogue.} The lower mass limit is driven not just by the $r$-band magnitude selection for GAMA, but also the need for each source to have a  calculated stellar mass. As this is determined using a multi-band $ugriz$ plus VISTA VIKING \citep{Edge2013} near-IR SED, fitted across the restframe range 3000-11000 \AA~ \citep[see][]{Taylor2011}, including the effects of dust extinction (using a \cite{Calzetti2000} extinction curve), a detection is required in several bands.  The VIKING observations replaced the UKIDSS near-IR data originally discussed, but not ultimately used, in \cite{Taylor2011}, the newer data giving consistent results across the full wavelength range. These stellar masses are also entirely consistent with masses derived from {\it WISE} data in the mid-IR  \citep{Cluver2014,Kettlety2017}.
 
We include only systems with $r$-band derived axial ratios $b/a>0.5$ in order to exclude those with edge-on disks which might provide  less reliable intrinsic photometry, given the large and potentially uncertain dust corrections made necessary by the strong obscuration presented by the disk. \citep[The adopted single value, or `screen' model, of the dust extinction will likely be least accurate for stars in edge-on disks;][]{Driver2007}. The mass, redshift and orientation selection leads to a volume complete sample of 12625 galaxies (split into three colour subsamples: 4455 blue, 2298 green  and  5872 red, see below).

In addition to providing a complete volume limited sample, our stellar mass selection is also driven by the need for the mass range to contain significant populations of galaxies over the full range of stellar population colours at these redshifts if we are to have any chance in identifying a population transitioning across the green valley. At higher stellar masses, the galaxy population is dominated by passively evolving red galaxies, while at lower masses the blue star formers dominate (see Fig. \ref{CMDa} and discussion in \citealt{Taylor2011,Taylor2015}). So at these mass extremes, in the redshift range covered, it would appear that any transition is either yet to begin (low mass) - if it ever will - or is completed (high mass), or in the case of the most massive galaxies has a very different evolutionary history \citep{Vandokkum2013}. Consequently at $z<0.2$ the chosen mass range will contain objects undergoing photometric transition, while it is by no means clear that this is the case for higher or lower masses. Obviously, this mass range encompasses the well-known point at which the low redshift galaxy population switches between being dominated by early- and late-type galaxies \citep{Kauffmann2003}.

Note that by choosing a fixed mass range across all colours, in particular when interpreting the results, we are implicitly assuming that galaxies do not evolve significantly in stellar mass while crossing the green valley (the same assumption as made by \citealt{Fang2013}). Mergers are known to be rare at low redshifts and are found to produce insignificant mass growth ($\sim 0.02$~dex/Gyr or less) at recent epochs \citep{DePropris2014}. Further, our green valley galaxies typically have SSFR $\sim 10^{-10.5}$yr$^{-1}$, i.e. a stellar mass increase $\sim 0.015$~dex/Gyr. Over 1-2~Gyr (see below), these are very small changes relative to our bin width of 0.5~dex, and to the expected mass errors of up to 0.1~dex \citep{Taylor2011}.

We split this sample into three  broad colour bins based on the surface density of points in the colour-mass plane.  For this we chose to use the intrinsic $u^*-r^*$ colour of the stellar population, i.e. the rest-frame colour of the best-fitting spectral energy distribution having removed the effect of internal and external reddening \citep{Taylor2011} as this measure is clearly sensitive to the true evolutionary state of the stellar population \citep{Taylor2015}. We note here that the definition of the `green valley' is somewhat nebulous, depending on the bands used and how well dust has been accounted for.  The original definition came from GALEX observations discussed by \cite{Wyder2007}, \cite{Martin2007} and \cite{Salim2007}, amongst others. Here we take advantage of the careful determination of the intrinsic stellar SED carried out in the optical for the GAMA sources \citep{Taylor2011}, which is made possible by the multi-band photometry and complete spectroscopy of the sample. As we are searching for evidence of truly transitioning populations with a global decline in star formation rate, rather than short-term variation up or down in SFR, these intrinsic optical colours are exactly what is required as they allow us to identify objects with stellar populations intermediate between purely passive and those undergoing so-called ``main sequence" star formation \citep{Noeske2007, Speagle2014} which dominate in the blue cloud.

This intrinsic colour  correlates  with the specific star formation rate (i.e. star formation rate per unit stellar mass; SSFR). The GAMA database provides multiple derivations of this statistic, from simple conversion of  optical line luminosities (as measured through a $2''$ spectroscopic fibre) to measurements on the complete SED of each object \citep[see the summary in][]{Davies2016}.  For Fig. \ref{SSFR} we show the measure
of specific star formation rate derived from a MAGPHYS \citep{Cunha2008}  analysis
of the 21-band far-UV to mid-IR SED of each object (Driver et al.
2016; Wright et al. 2016), relatively insensitive to stochastic variation on the shortest timescales. Other measures give equivalent results.

 We note that others have studied similar or lower redshift SDSS galaxies as a function of rest-frame colour without being able to correct for internal reddening \citep[e.g.][]{Baldry2004, SanchezAlmeida2010, Casado2015}.  Inevitably, not only does this make the colours of the entire population redder, but it tends to scatter objects with intrinsically blue stellar populations into any green valley defined using uncorrected colours while intrinsically green objects may be moved into  the red sequence. The use of the intrinsic stellar colours afforded by the GAMA analysis is therefore  advantageous in studying evolution of the underlying stellar population across the green valley. 

We classify galaxies as red if they are redder  than $u^*-r^* = 1.5$ at log($M_*/M_{\odot}$) = 10, rising linearly with log($M_*$) to  $u^*-r^* = 1.7$ at log($M_*/M_{\odot}$) = 11, though only selecting galaxies in the  mass range noted above. This selects all the objects in the red sequence over the required mass range.  Blue galaxies are defined as those bluer than a line similarly connecting $u^*-r^* =1.3$ to 1.4, again covering the parameter space where sources typically classified as blue cloud galaxies reside. Galaxies are denoted as green between these two lines. The midpoint of these two lines  follows the minimum surface density of objects in this space as a function of mass, essentially the centre of the `green valley' as normally defined, at least in these bands.   While these cuts can be considered somewhat arbitrary and may not select out exactly the same objects as when a selection is made using a different colour combination (e.g. NUV$-r$), as noted above there is no single definition as to what constitutes a green valley galaxy \citep[e.g. see][]{Salim2014}. In any event, in this paper we consider galaxy properties across a continuum of intrinsic colours and use the three broad colour bins for largely descriptive purposes, a similar approach to that used in a recent paper by \cite{Oemler2017}, albeit in that case galaxies were divided into three samples based on cuts in the continuum of their SSFR values. 

\begin{figure}
\includegraphics[width=\linewidth]{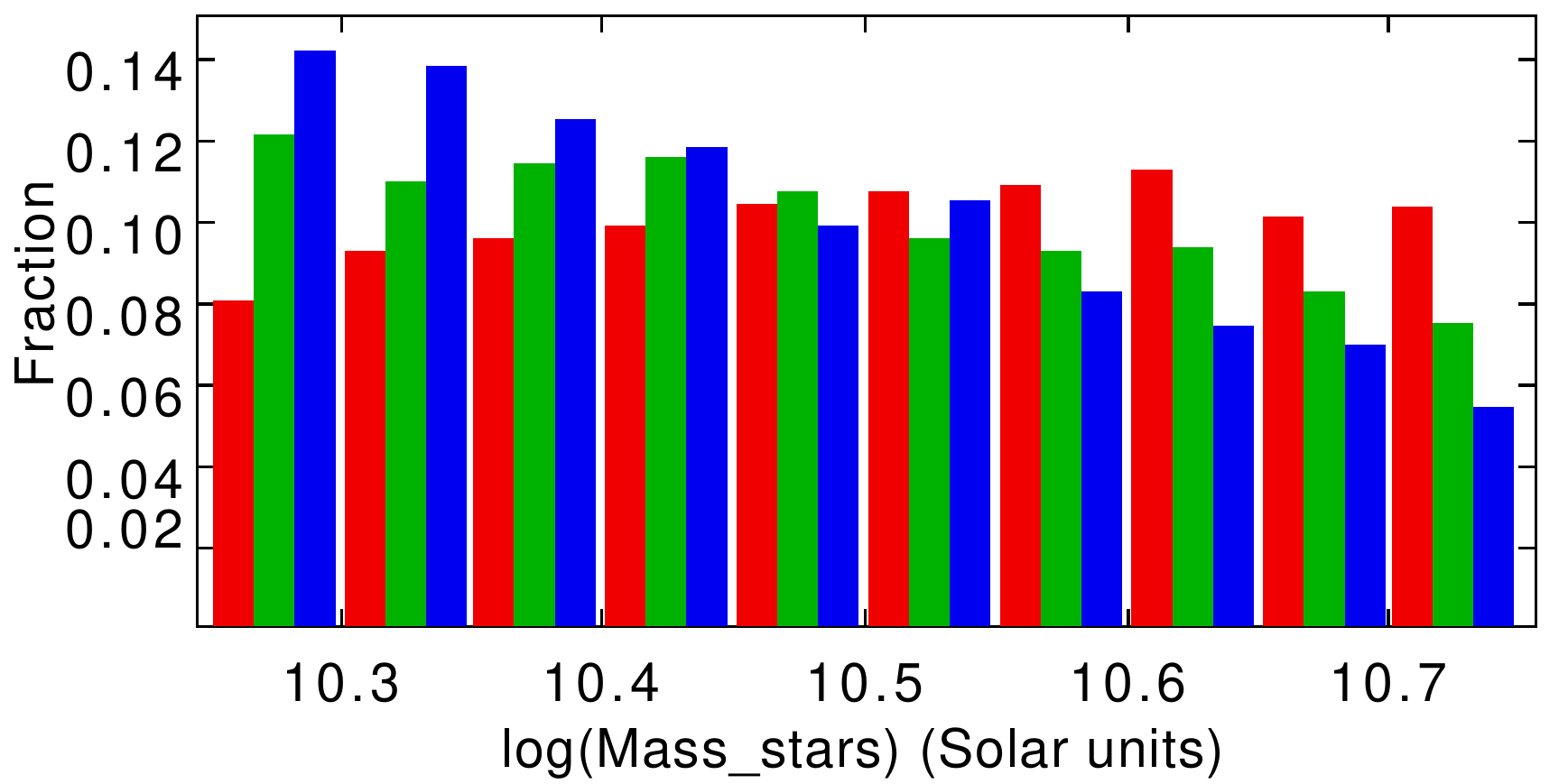}
\includegraphics[width=\linewidth]{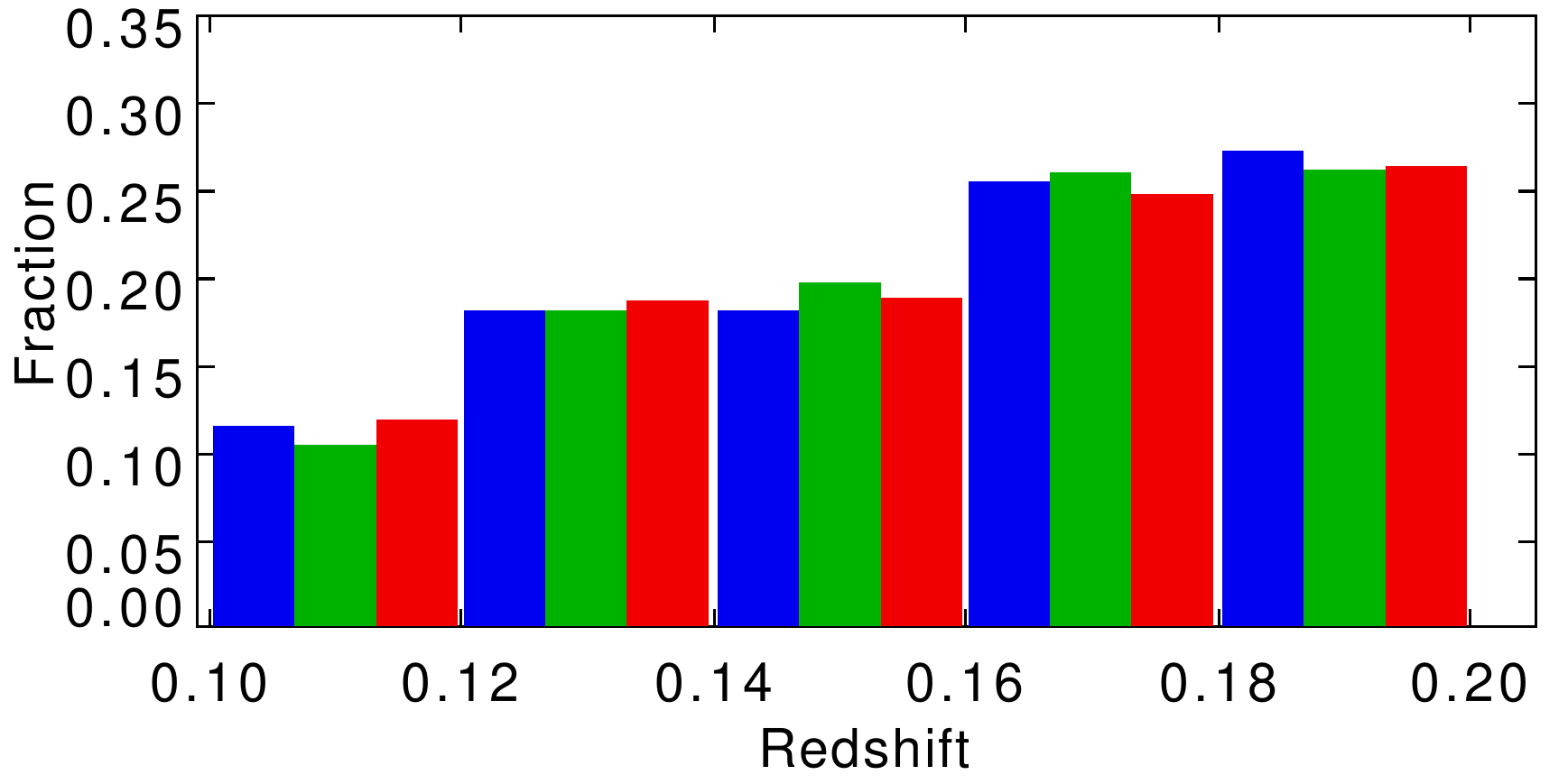}

\caption{Top: Stellar mass distribution within the three subsamples (coloured appropriately), bin width is 0.05 dex in mass.  As expected, blue galaxies are somewhat more prevalent at the low mass end. Bottom: Redshift distribution within the three subsamples, bin width  is  0.02 in redshift.}
\label{mass_z}
\end{figure}

The majority of the green sample have MAGPHYS-determined SSFR averaged over the last Gyr of 
$10^{-10.8}<  {\rm SSFR}<10^{-10.2}$yr$^{-1}$ with a tail to lower values. Most blue galaxies have   significantly higher SSFRs (to ${\rm SSFR} \sim10^{-9.5}$yr$^{-1}$) over the same time period and the red galaxies typically have  ${\rm SSFR}<10^{-10.7}$yr$^{-1}$ and often significantly lower  (see Fig. \ref{SSFR}). Similar results are obtained for the SSFR averaged over the last 100 Myr, albeit with a slight shift to lower SSFR for mean values and an increase in  range due  to higher stochasticity over the shorter time period. See \cite{Wright2016} for details of the application of MAGPHYS to the GAMA survey.  The redshift distributions of all three subsamples are essentially identical (Fig. \ref{mass_z}). The mass distribution varies between subsamples (Fig. \ref{mass_z}) exactly as expected, with the blue fraction increasing slightly towards lower masses.

\section{Structural Analysis}

\begin{figure}
\includegraphics[width=\linewidth]{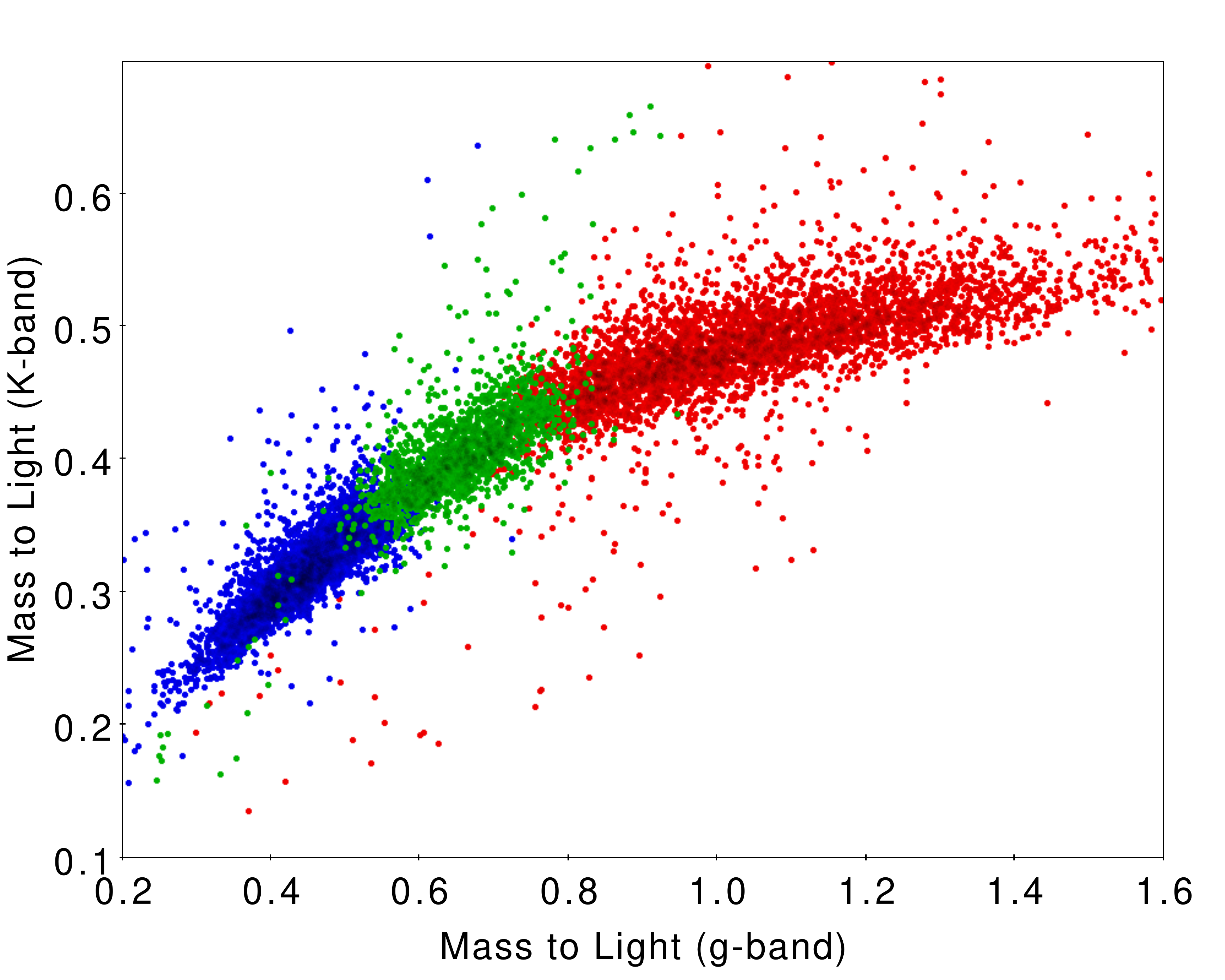}

\caption{Mass--to-light ratios in Solar units  measured in the $g-$band (x-axis) and $K-$band (y-axis) for our three subsamples, plotted colours as before.  It is evident that $M_*/L$ varies far more with galaxy colour in $g$ than in $K$ as expected. The difference in $K-$band $M_*/L$ between the  typical red and green galaxies is of order 0.1 dex  while the difference in the same quantity between the the blue and the red subsamples is greater than 0.2 in dex.  As expected, the $g-$band mass-to-light ratio strongly tracks the $u^*-r^*$ colours of the galaxies, the bluest having the lowest ratios, reddest having the highest.}

\label{moverl}
\end{figure}

Having selected this primary sample and split it into three colour-based subsamples, we can now explore the structural parameters of the individual galaxies. In order to do this,  we use a standard GAMA data product available for each object, a single component S\'ersic fit  which is determined for each band from $u$ through to $K$. In the following, we use the results in both the $g-$ and $K-$bands. The $g-$band results represent the observed optical structures, which depend on both the underlying (radial) stellar mass distribution for each galaxy and the radial variation  in mass-to-light ratio in the $g-$band. Given that the star formation histories of bulges and disks in individual galaxies are likely to be different, optical mass-to-light ratio variation may well have a significant effect on perceived morphology. However, the $K-$ band measurements will give  a significantly cleaner measure of the underlying stellar mass distribution for individual galaxies. While there is still some variation in $K-$band mass-to-light ratio with stellar population age (a factor of 2 between the extremes of the blue and red subsamples), it is much smaller than that at optical wavelengths (a factor of $\sim 7-8$) , as shown in Fig. \ref{moverl} \citep[see also][]{Bell2003}. 

This part of the analysis does not allow direct bulge/disk separation. However, the objectively measured S\'ersic index $n$ does accurately characterise the radial distribution of $K-$band light (and therefore  approximately stellar mass) at these redshifts (in essence, $n$ will increase with mass concentration and therefore bulge prominence). Thus we are able to demonstrate any variations in the mass profile as a function of intrinsic colour and from subsample to subsample.

The $K-$band data for each object are taken from VIKING, the VISTA Kilo-Degree Infrared Galaxy Survey \citep{Edge2013}. This produces images with a $5\sigma$ depth in a 2-arcsec aperture  of $K_s\sim 21.2$  where $K_s$ is the VISTA short $K-$band filter. Typical seeing is 0.9-1.0  arcsec (see \citealt{Venemans2013} for further description of the survey). The fits provide parameterisation in terms of S\'ersic index, effective radius and apparent ellipticity of each object. The fit uses a psf appropriate to each object in each band derived from point sources close to the object in question. The S\'ersic fit is carried out using {\sc GALFIT 3} \citep{PengCY2010}. Details of the process as applied to GAMA are given in \cite{Kelvin2012}. 

While the $K-$band light distribution is a closer match to the stellar mass distribution than is the optical light distribution, there is still a component of near-IR light that arises from younger populations. This is reflected in a small variation in mass to $K-$ band light ratio with optical colour - with typical blue objects having a lower $M_*/L_K$ than red objects  by a factor of at least 0.2 dex (see Fig. \ref{moverl}).   Consequently, while the similarity in $M_*/L_K$  between the red and green samples means that in both cases the light can be assumed to  trace mass sufficiently accurately for our purposes, this is not exactly true for the blue subsample, something that must be borne in mind when interpreting differences in S\'ersic fits between the subsamples. We note there is no band which perfectly traces stellar mass, even $M_*/L$ at 3.5 $\mu$m  has a dependence on optical colour \citep{Cluver2014}, but for our purposes we can interpret the $K-$band light distribution as indicative of the radial stellar mass profile subject to this {\it caveat}.  

\subsection{Photometric parameters}
\label{phot}

\begin{figure}

\includegraphics[width=\linewidth]{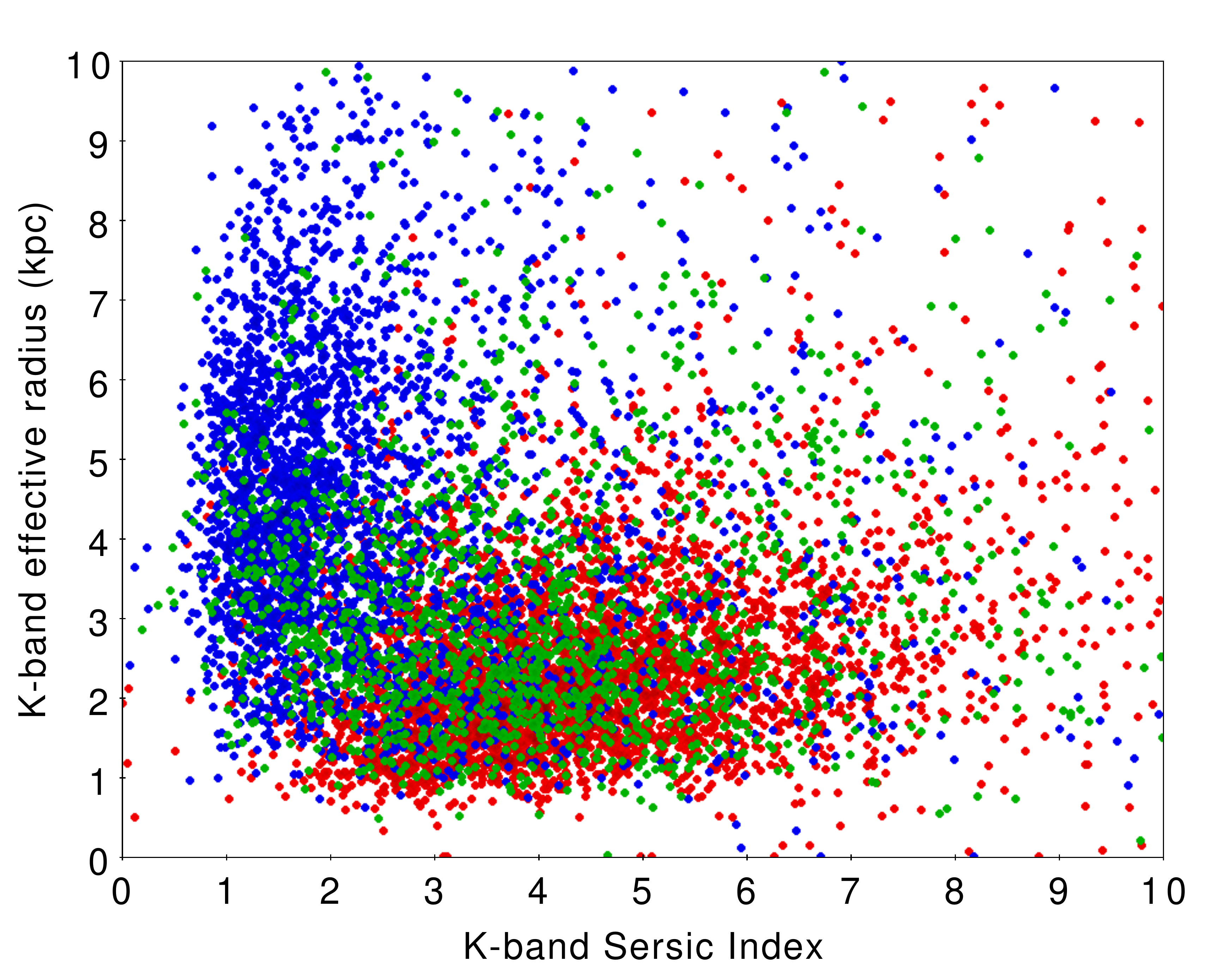}
\includegraphics[width=\linewidth]{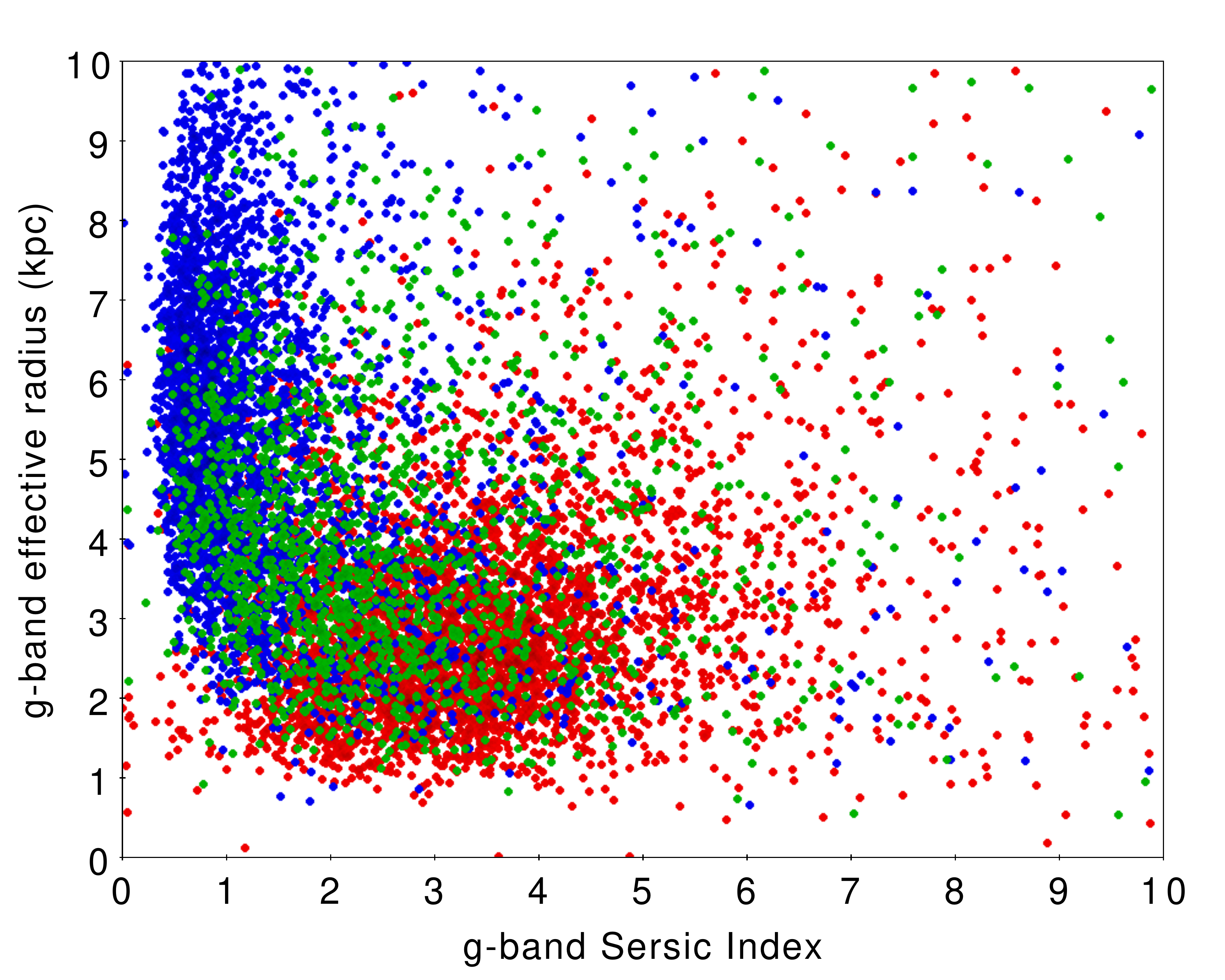}

\caption{ Top: K-band S\'ersic index - effective radius plane  for our red, green and blue subsamples, colour-coded appropriately. As expected the red and blue subsamples generally inhabit different regions within this space: the red objects being spread more towards higher S\'ersic indices and smaller effective radii, as appropriate for an early type, spheroid-dominated population, while the blue systems are typically more extended and consistent with the near exponential profiles of a disk dominated population. The parameters of the green population are a much closer match to those of the red than the blue population. Bottom: The same plot, but this time with the structural parameters measured in the $g-$band. The blue galaxies are seen to be even more extended and disk like ($n \sim 1$), while the green galaxies are now intermediate between the reds and blues. Fractional uncertainties on individual points  are 0.2 and 0.15 for $R_e$ and $n$ in the $K-$band and 0.1 and 0.15 in the $g-$band.}
\label{nvre}
\end{figure}

A plot of the $K-$band S\'ersic index - effective radius plane is shown in Fig. \ref{nvre} (upper panel) for the three subsamples.
Given that the (intrinsic) effective radius of an object is determined by convolving the S\'ersic model with the appropriate PSF for that object, the lower limit to this parameter ($\sim 0.4-0.5$ arcsec, corresponding to $\sim 1.5$~kpc at our median $z$) is essentially set by the VIKING seeing. Any galaxies smaller than this would have only an upper limit to their intrinsic $R_e$ with our data. The plot shows the results for 92 per cent of the total sample; the fits to the rest result in implausible values for the parameters or fail to give a reasonable fit at all. The missing galaxies are spread uniformly over the three colour subsamples.  It is immediately apparent that the red and blue subsamples follow our preconceptions, namely the red galaxies are compact and generally have S\'ersic indices $n>2.5$ indicating spheroid-dominated systems and the blue galaxies are almost all $n<2.5$ systems dominated by an extended disk component \citep[cf.][]{Brennan2017}, though we should note that there is a fraction of the blue population whose parameters extend into  the region occupied by red galaxies \cite[cf.][]{Taylor2015}.  The green sources largely fall in the same region of parameter space as the red galaxies, though again with an overlap into the `blue' dominated region.  The fractional uncertainty on an individual $R_e$ is typically 0.2 and on $n$ it is 0.15. 

This behaviour can be contrasted with that seen when plotting the same parameters as measured in the $g-$band (lower panel of Fig. \ref{nvre}).  Here we see that the green subsample is, as expected, intermediate between the red and the blue.   Evidently, in terms of radial {\em mass} distribution, green and red galaxies are extremely similar, but this concordance is hidden when we use (blueish) light profiles.  Removing the relatively small variation in $M_*/L_K$ between the green and red subsamples  noted in the previous subsection would only act to make their typical radial mass distributions  even more similar than indicated in the top panel of Fig. \ref{nvre}.
 The fractional uncertainty on an individual $g-$band $R_e$ is typically 0.1 and on $n$ is 0.15. 

\begin{figure}

\includegraphics[width=\linewidth]{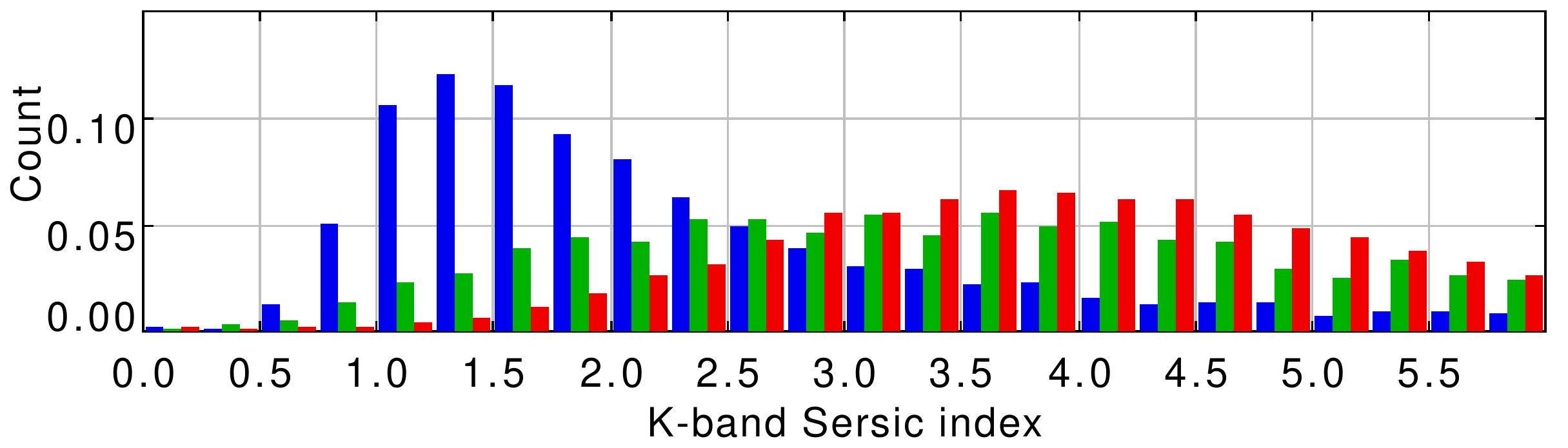}
\includegraphics[width=\linewidth]{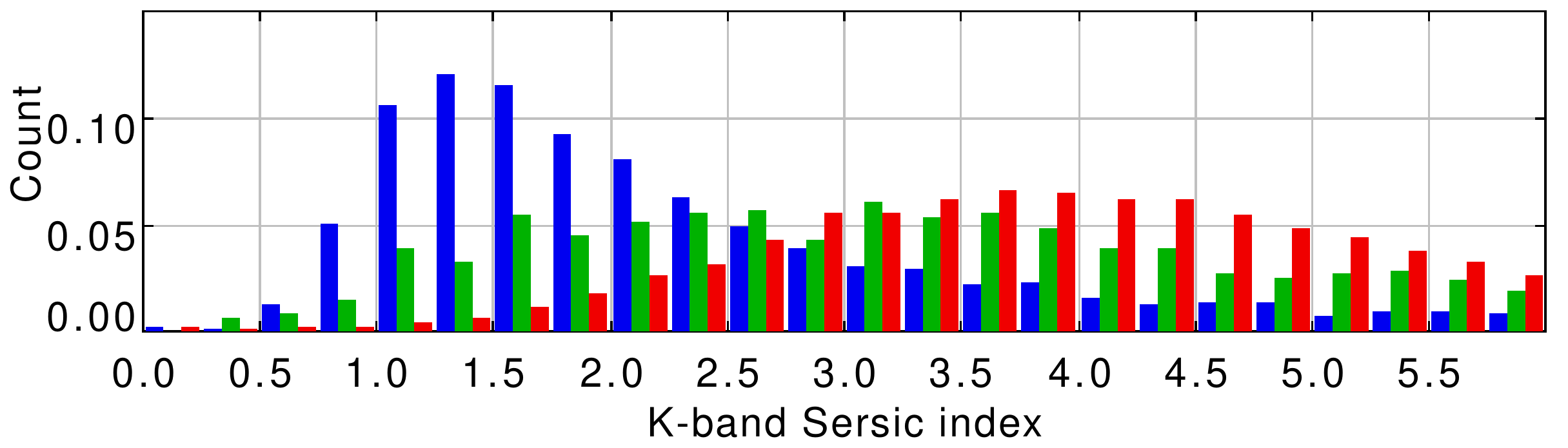}
\includegraphics[width=\linewidth]{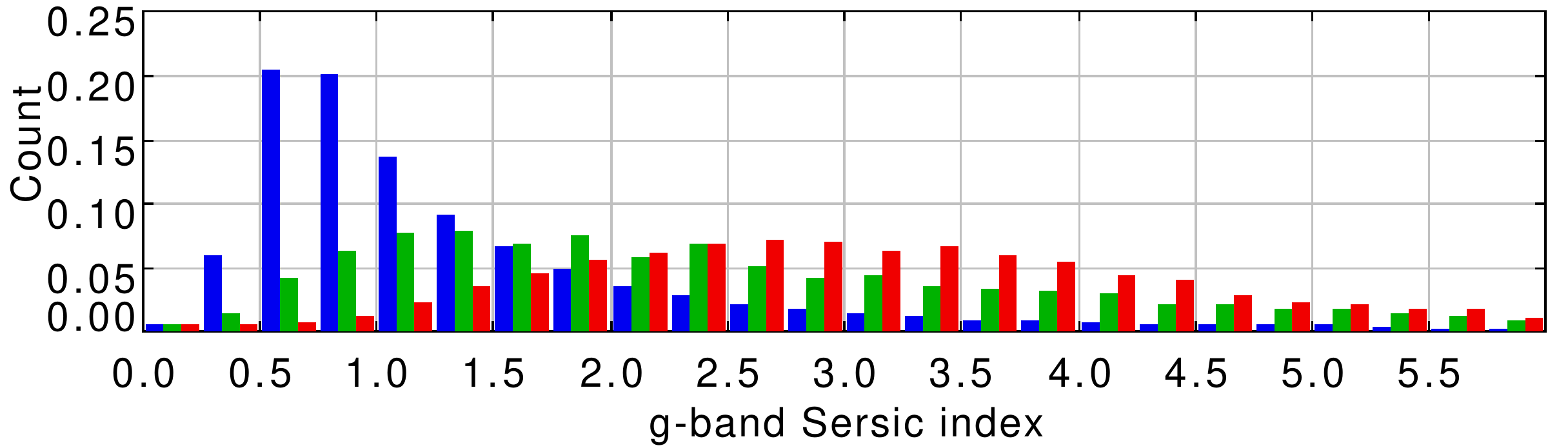}

\caption{ Top: Normalised histogram of $K-$band S\'ersic index distribution for the red, green and blue subsamples, clearly demonstrating the significant difference of the distribution for the blue galaxies  and the similarity for those of the green and red galaxies. Less than 10 per cent of the green galaxies need to be re-distributed in S\'ersic index for the red and green distributions to be statistically indistinguishable. Middle: The same plot, but this time only including the bluest third of the green sample. These have a mean $u^*-r^*\sim 1.35$ , only 0.15 magnitudes redder than the mean colour of the blue subsample. Despite this small difference in colour, the green distribution is still far closer to that of  the red subsample (which has a mean $u^*-r^*\sim1.9$) than that of the blue, with only a $\sim 5$ per cent increase in the fraction of sources with $n<2.5$ in comparison to the complete green sample. Bottom: The same as the top histogram, but this time measuring the distribution of $g-$band S\'ersic indices. In this case the complete green subsample has a S\'ersic distribution intermediate between those of the blue and red subsamples, as expected from previous work. The difference in $g-$ and $K-$band measurements are caused by the near-IR surface brightness profile being a better tracer of the underlying stellar mass distribution than the optical surface brightness profile and thus demonstrating that the mass profiles for green and red galaxies are extremely similar. }
\label{hist_sersic}
\end{figure}

We further explore this by plotting the histogram of  the S\'ersic index distribution for all subsamples, and additionally a subset of the bluest third of the green subsample (see Fig. \ref{hist_sersic}). From this, it is clear that the distribution for the green galaxies is far closer to that of the red systems than that of the blue. This remains true when restricting the green subsample to the bluest third of members. These systems have a mean intrinsic $u^*-r^*$ colour of their stellar populations only $\sim 0.15$ magnitudes above the same mean quantity for the blue galaxies but  0.65 magnitudes bluer  than that for the red galaxies.  While the fraction of  sources with $n<2.5$ is increased by of order 5\% relative to the whole green subsample, the shape of the rest of the distribution is hardly changed despite the much closer proximity in colour space of this green subsample to the blue galaxies then the red ones. 

 Looking back at Fig. \ref{SSFR} this behaviour could be considered somewhat surprising because at the blue-green boundary in this plot there is significant overlap in the SSFRs of the galaxies at either side. To further explore this,
we compared the same morphological parameters  for blue galaxies with SSFR$<10^{-10.3}$ and green galaxies with SSFRs above this value (so these green galaxies are apparently more active by this measure than the blue).  Even in this case, these blue and green subsamples differ in the structural  parameters  in the same way as the full blue and green samples. Evidently, for the measurements used in this paper, the intrinsic colour of the stellar population more closely correlates with the overall structure of the galaxy than does the estimated star formation rate.

All of this indicates that any change in underlying structure as measured in the $K-$band (so most closely following the radial mass distribution of the stellar population)  sets in  at the blue cloud/green valley boundary, rather than smoothly varying across the range of stellar population colours. In turn this suggests that a bulge dominated mass distribution is required {\em before} a galaxy can become green \citep[cf.][]{Cheung2012}.  \citealt[][]{Oemler2017} (their Fig. 11) similarly find that the distribution of bulge mass fractions of their truly passive and ``quiescent" samples (objects with SSFRs significantly lower than those of blue cloud galaxies) are very similar and significantly different from those of star formers \citep[see also][]{Carollo2016}.  This behaviour seems to be in agreement with the theoretical prediction from semi-analytic models, that the median $n$ rises steeply  immediately below the simulated star formation main sequence \citep[Fig. 9]{Brennan2015}. 

\subsection{Environment}
 \label{env} 
 
It has been known since the earliest days of extragalactic astronomy that environment and morphology are linked \citep{HubbleHumason1931} and this was encapsulated in Dressler's (1980) morphology-density relation. As an additional constraint on the ways in which galaxies can traverse in colour from blue to red, we therefore next explore the environments in which green galaxies occur, with respect to red or blue galaxies. For this we use the GAMA environment parameter $\Sigma_5$, the projected local density within the distance to the given galaxy's fifth nearest neighbour within the GAMA sample with redshift difference less than 1000~km~s$^{-1}$ \citep[see][]{Brough2013}. We also make use of the `friends of friends' based GAMA Galaxy Group catalogue (G$^3$C) of \cite{Robotham2011}, as updated to cover all three equatorial fields in GAMA II.

\begin{figure}
\includegraphics[width=\linewidth]{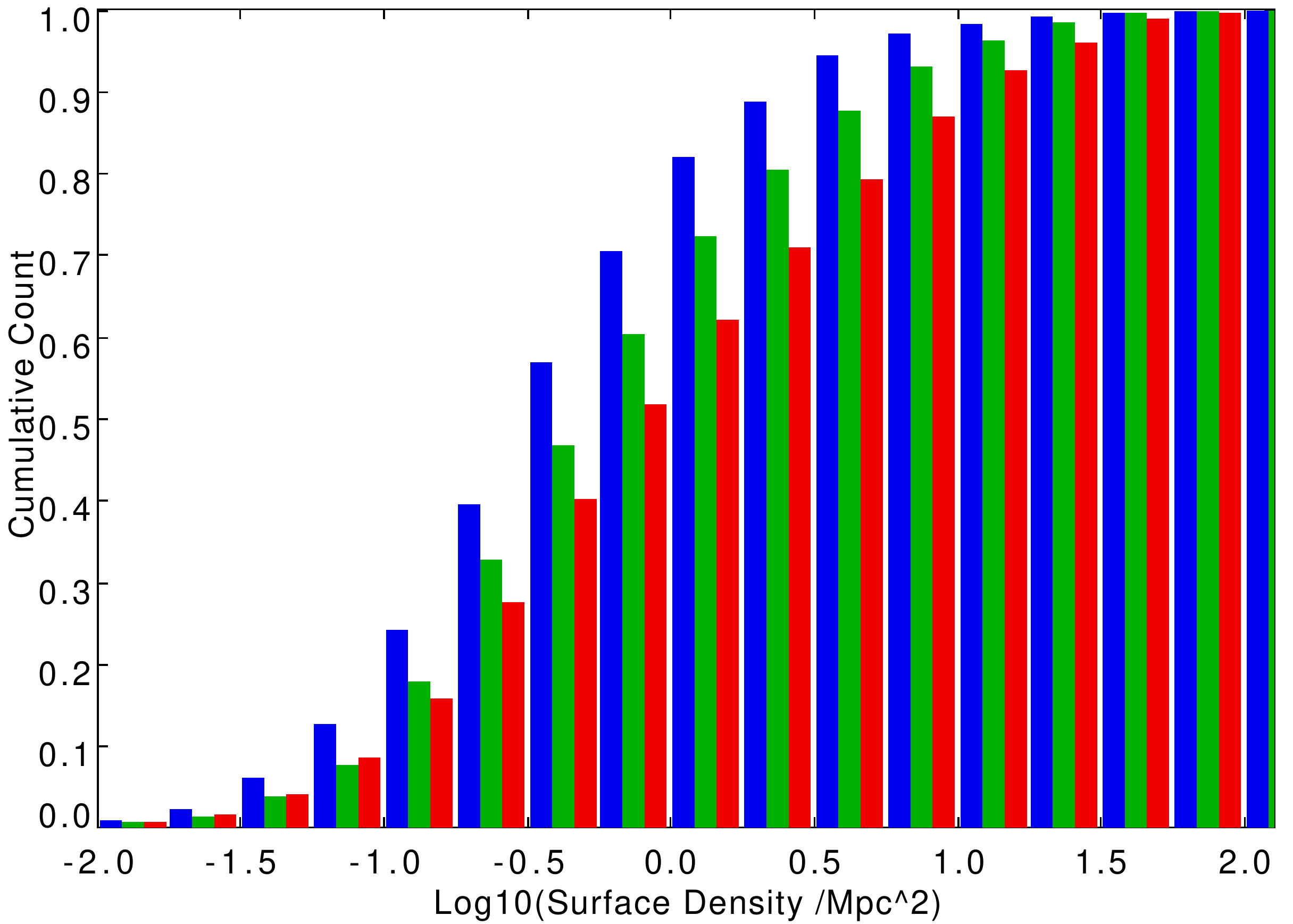}

\caption{Cumulative distribution of local environmental richness as measured by the surface density of  the 5 nearest neighbours within 1000 km/s for the three subsamples.  The typical local environment of the green subsample is clearly intermediate in richness between the others, {\it i.e.} richer than those of the blue galaxies but poorer than those of the red galaxies  }
\label{sigma5_hist}
\end{figure}

We can quantify environmental effects by plotting the cumulative distribution of local surface densities as measured by $\Sigma_5$ for the galaxies in the red/green/blue subsets (Fig. \ref{sigma5_hist}). This confirms that, at the 50th percentile of each distribution, the green galaxies live in regions of 0.2~dex higher density than blue galaxies, but 0.1~dex  lower density than red galaxies, generally avoiding both density extremes \citep[cf.][]{Bait2017}.  This may suggest that (blue) galaxies in the lowest density environments have not (yet) had the opportunity to turn green, whilst most (red) galaxies in the highest density regions have already passed through any green phase at higher redshifts.  

Notice that this is not exactly the same as the standard \cite{Dressler1980} variation in the fraction of S0 galaxies - arguably identifiable with green galaxies in some scenarios - since the S0 fraction increases monotonically with density, with their numbers only being overtaken by ellipticals at the very highest densities. In the present work classical S0s are generally found in the red subsample (see Table 2, below), but note that the GAMA equatorial regions contain no rich clusters in our redshift range.

We next consider the global environment of galaxies in groups. Here we use  G$^3$C to compare the ratios of blue to green to red galaxies as a function of group multiplicity.  In Table \ref{tab_env} we show the fraction of red, green and blue galaxies in our sample broken down by environment in terms of group multiplicity (counting only objects within the GAMA magnitude-limited catalogue). As expected the fraction of galaxies in our sample which are red increases with multiplicity and the blue fraction drops correspondingly. Note that we do not distinguish between `central' and satellite galaxies \citep[e.g.][]{Tal2014, Bluck2016}, but clearly all isolated galaxies, for instance, are centrals while most of those in large groups are satellites, even at our moderately high mass range. In all environments, the fraction of galaxies drawn from our sample that we classify as green remains approximately constant at between 15 and 20 per cent. \cite{Weinmann2006} previously found the same result for `intermediate' type (by colour and SSFR) SDSS galaxies in groups \cite[see also][particularly their table 7]{Coenda2017,Bait2017}, while \cite{vonderLinden2010} obtained a corresponding result for `green' galaxies (selected via a principal  analysis), as a function of position in clusters.

Given how the red and blue subsamples behave, the approximately constant green fraction likely depends on a number of factors, but  it must say something about the typical timescale of transition from blue to red in terms of the total lifetime of any system.  Also included in the table is the ratio of the number of green to blue galaxies.  If this can be interpreted as the ratio of their typical lifetimes in each of the environments, it indicates that typical  blue lifetimes decrease with increasing richness, clearly pointing to a role for environment in decreasing their star formation rates, as expected from many previous studies.

\begin{table}

\begin{tabular}{lccccccc}
Multiplicity & $N_R$&$N_G$&$N_B$&$f_R$&$f_G$&$f_B$ &$ N_G/N_B$\\
\hline
Field & 1774 & 810 & 1833 &0.40 & 0.18&0.42 & 0.44\\
2 to 5& 1332&533 & 1019& 0.46&0.18&0.36 & 0.52\\ 
6 to 10&334&106 & 143&0.57& 0.18&0.25 & 0.74\\
11 to 20 &252&80&73&0.62& 0.20&0.18& 1.10\\
21+ & 505&105&79 & 0.73& 0.15 & 0.12& 1.33\\
\end{tabular}
\caption{Number and fraction of total sample for red, green and blue subsamples as a function of environment, as measured by multiplicity of the group within which each galaxy is found. Multiplicity is determined from the magnitude-limited GAMA catalogue, see \citet{Robotham2011}. The green fraction is independent of multiplicity, while there is the expected shift of blue to red colours with increasing richness. }
\label{tab_env}
\end{table}

Going beyond the colours themselves, it is potentially of interest to look at the distributions of structural parameters in the different  environments. Having compared $g-$ and $K-$band effective radii and S\'ersic indices for the three colour subsamples as a function of environmental multiplicity and local surface density, we identified no statistically-significant differences in the distributions. This demonstrates that the band-dependent differences in morphology between the three subsamples, and particularly those between the blue and green subsamples persist across environments. This implies that even if there is an environmental component to triggering a transition from blue to green, as implied by the environmentally-dependent blue fraction, once started the transition process is broadly the same in any environment. Alternatively, different environmentally-dependent processes have the same timescales and eventual results, either individually or when acting in combination \citep[see e.g.][]{Smethurst2017}.

\section{Morphologies}
\label{morpho}

\subsection{Bulge-Disk Decomposition}
\label{BDD}

Having carried out the single-component S\'ersic analysis described above, we can explore the morphologies of galaxies within the colour subsamples by using the results of a previously completed bulge-disk decomposition carried out on a subset of our objects. \cite{Kennedy2016} presented the results of bulge-disk decompositions for $z<0.3$ galaxies drawn from a subset of the area of one of the equatorial fields (G09).  A two component fit was made to the surface brightness profile of each object, with one component having a fixed S\'ersic index, $n=1$, so assumed to represent a disk, whilst the S\'ersic index of the other component was free. The effective radii of the bulges and disks, and the components' luminosities, were left free.  Profile fits were made simultaneously at each wavelength, $u$ to $K$, constrained such that the individual components' profile parameters $n$ and $r_e$ were constant with wavelength, using the methods presented in \cite{Haeussler2013}. Physically, the relative normalisation of the components at different wavelengths reflects their relative mass-to-light ratios in each band.  Here we select objects from our main sample which were also analysed by \cite{Kennedy2016}.

 We firstly use their results to determine the $K-$band absolute magnitudes of the spheroid components of our red, green and blue subsamples. Depending upon whether an object has two robust components in its fit, we determine this magnitude in two different ways. For those systems where the radial profile is in fact fitted better by a single component {\it and} the  single $K-$band S\'ersic index is $n>2$,  we use the catalogued total absolute  rest-frame dust-corrected $K-$band magnitude as determined by the methods used in \cite{Taylor2011}. For those sources where both bulge and disk are deemed present and the bulge is no more than three magnitudes fainter than the disk in $K$, and has a S\'ersic index of $n>2$, we use the same absolute magnitude scaled by the bulge fraction.  We estimate the error on any individual determination to be $\sim 0.3$ magnitudes.

\begin{figure}
\includegraphics[width=\linewidth]{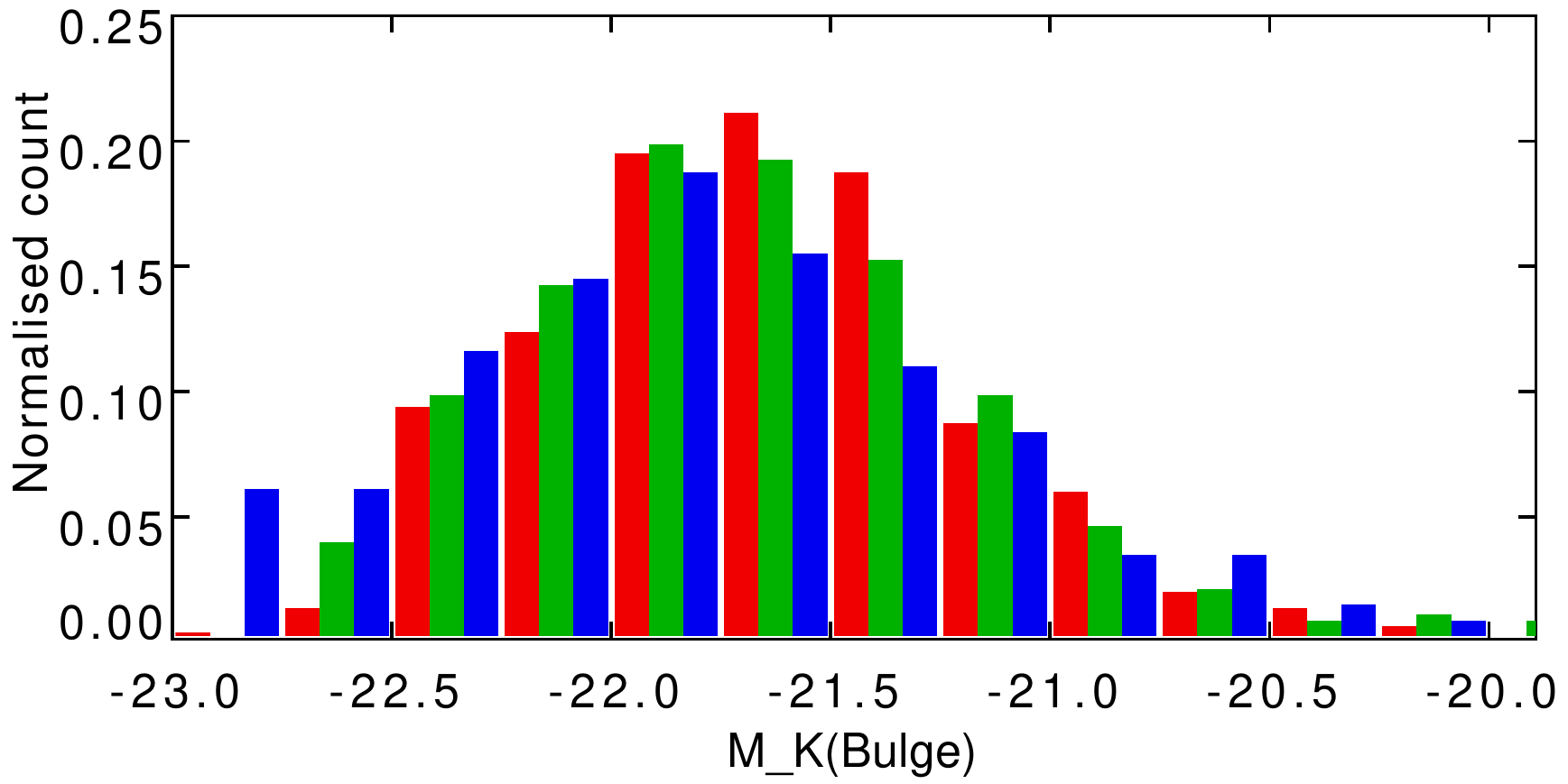}

\caption{Distribution  of absolute $K-$band magnitudes for the bulge components (of systems with a good bulge-disk decomposition)   plus those objects best fit as dominant spheroids from their single component S\'ersic fit.  The distribution of magnitudes for the red, green and blue subsamples are comparable, as are the fraction of red and green galaxies contributing to this plot (70-80\% of each sample), so the distribution of $K-$band luminosities and therefore masses for the bulges of the red and green subsamples in this mass and redshift range are comparable. Only a minority (40\%) of the blue galaxies show a separable bulge component and are included in this plot, indicating that a subset of the blue galaxies have bulge  masses and $K-$band luminosities comparable to those in the green and red subsamples.\label{bulge_mag}
}

\end{figure}

The histogram of bulge/spheroid absolute magnitudes is shown in Fig. \ref{bulge_mag}, broadly indicating that its distribution is indistinguishable between the subsamples: a Kolmogorov-Smirnov (KS) test returns no evidence for different distributions at the P=0.1 level.  Some care should be taken in the  interpretation of these distributions as they  only include those objects with plausible bulge components and do not take into account those galaxies with small or negligible bulges. Of the 2460 galaxies in our main sample drawn from this part of the  G09 field,  976 of 1231 red galaxies were included in this plot, as were 318 of 454 green and 305 of 760 blue galaxies. So for the red and green subsamples the statistics are  similar, consistent with the single S\'ersic analysis above. Blue galaxies with bulges strong enough to meet our criteria are in the minority as the majority of the blue galaxies have single S\'ersic fits with $n<2$.  Given the comparatively limited variation in $K-$band $M_*/L$ across the three subsamples (Fig. \ref{moverl}), the bulge mass distribution of the red and green subsamples is indistinguishable given the current data, as is that for the $\sim 40$ per cent of blue galaxies with the strongest bulges.
 
 An obvious implication of these results is that if the green galaxies in this redshift and mass range were previously blue star forming systems that subsequently reduced their specific star formation rate, they must have formed their bulges or otherwise re-arranged their radial stellar surface density prior to their entry into the green valley. Both this and the results in Section 3.1 and Fig. \ref{nvre} support and extend the finding by \cite{Fang2013} and others that green galaxies have similar central surface densities to those of red galaxies, while blue galaxies have a range of central surface densities extending to significantly lower values. Those blue galaxies with parameters overlapping those of the bulk of the green and red populations in Fig. \ref{nvre} are those blue objects with significant bulges and therefore the most likely immediate progenitors of the green galaxies  from within the blue population.

\begin{figure}
\includegraphics[width=\linewidth]{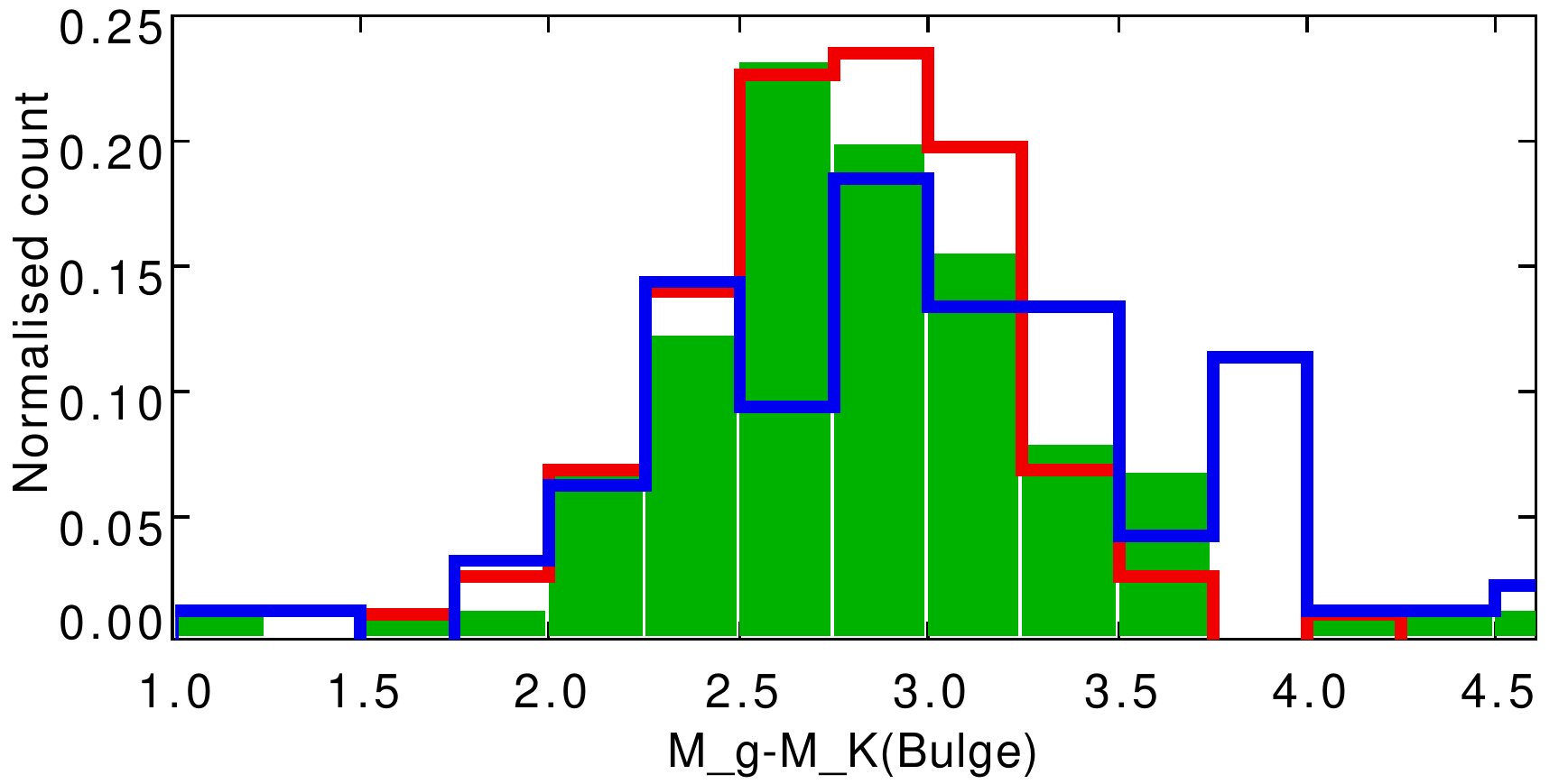}

\caption{Distribution of rest-frame $g-K$ colours for the bulge components in a subset of those galaxies identified as having both bulge and disk components in Kennedy et al., (2016), as detailed in the text. The distributions for the three colour subsamples are plotted in red, green and blue.}

\label{bulge_col}
\end{figure}

\begin{figure}
\includegraphics[width=\linewidth]{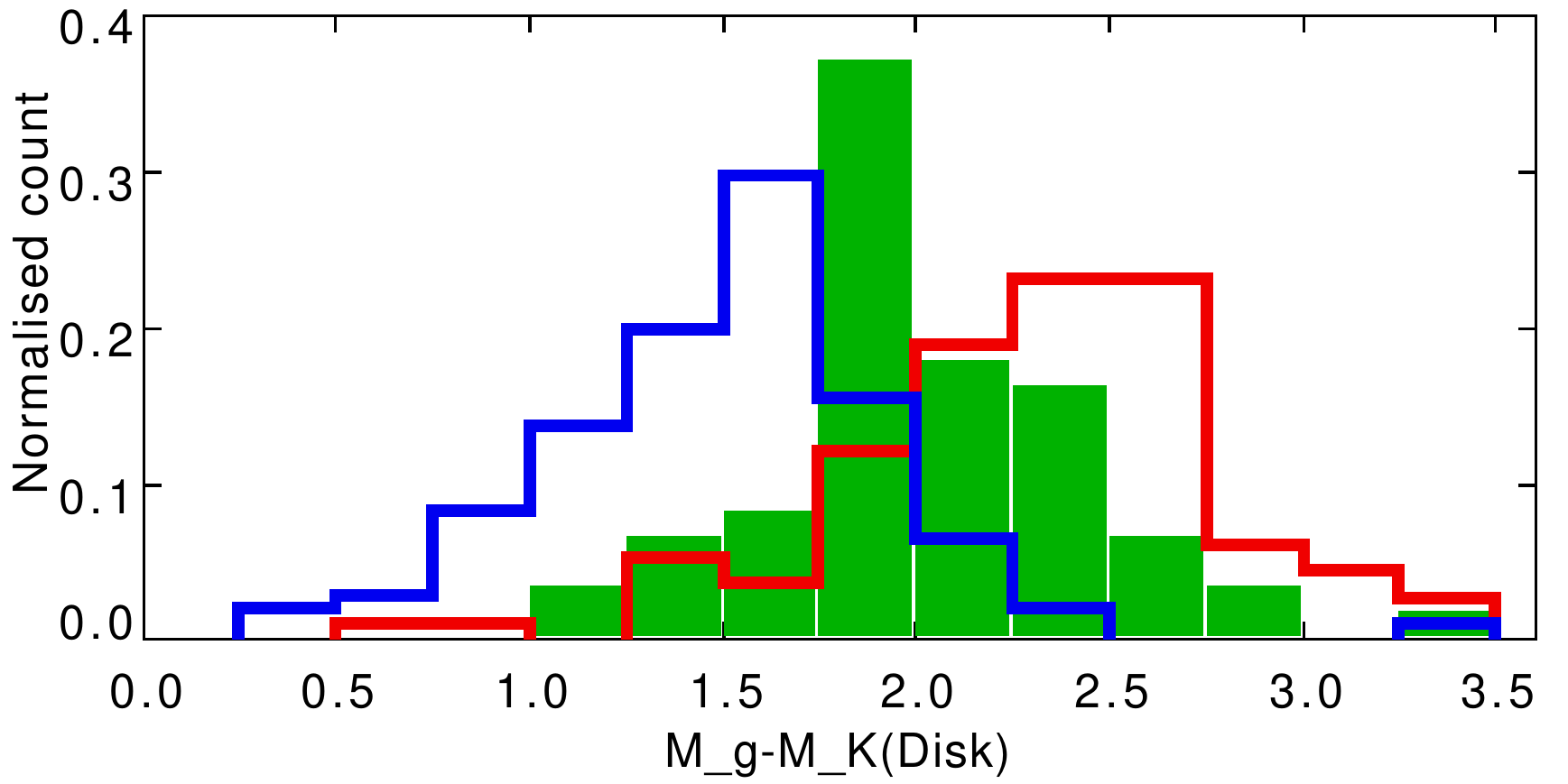}

\caption{Distribution of rest-frame $g-K$ colours for the disk components in a subset of those galaxies identified as having both bulge and disk components in Kennedy et. al., (2016), as detailed in the text. The distributions for the three colour subsamples are plotted in red, green and blue.}

\label{disk_col}
\end{figure}

 We use the same bulge-disk decomposition to explore the difference in bulge and disk colour  across the three subsamples in order to find clues as to which morphological component is driving the colour change from blue to red. In this instance we only use those objects where two components were identified and differed  in $K-$ band magnitude by less than 3 magnitudes (so one component did not completely dominate), and where the bulge component S\'ersic index is between $2<n<7.5$. The lower limit rejects components that are unlikely to be true bulges and the upper limit is set to avoid objects where the fit defaulted to a value of $n=8$  when the component was extremely centrally concentrated \citep{Kennedy2016}. We additionally require  an uncertainty in the magnitudes in both $g-$ and $K-$bands to be $<0.3$ mags for the colour of a bulge or disk to be included in the following analysis.   We plot histograms of the bulge and disk colours separately in Figs. \ref{bulge_col} and \ref{disk_col}.  These demonstrate that the distribution of bulge colours for all three subsamples is very similar (Fig. \ref{bulge_col}) while there is a clear trend in the disk colours with blue two-component galaxies having bluer disks than green two-component galaxies which in turn have bluer disks than the red two-component galaxies (Fig. \ref{disk_col}). The three disk colour distributions shift by about 0.5 magnitudes between each of the subsamples.  A KS test confirms that the three distributions differ at the P=0.001 level.  
 
 We should note that these colours are not corrected for differential dust reddening between the two components. The comparatively red colours ($g-K\sim 4$) of a small fraction($\sim 10$ per cent)  of the blue galaxy bulges may be due to the reddening of the bulge optical light by the central regions of their disks, as in \cite{Driver2007}. A KS test excluding these shows no difference between the bulge colours of the green population and that of either the blue or  the red at the P=0.1 level. Including these outliers leads to a difference between the red and blue populations at the P$\sim 0.01$ level, though not between the green and the blue populations.  
 
 These results imply that the colour change of the overall stellar SED is primarily driven by the colour of the disk in systems with clear bulge and disk components. Any possible difference in the bulge  colour distribution for the blue galaxies relative to those of the  other populations only works to strengthen this result, given  that the discrepant bulges are too red. This analysis is evidently supported by   the visual impression given by Figs. \ref{tci} and \ref{tci_low} (see below) where the bulges of galaxies in all three colour subsamples are rendered similarly, while the disks of the blue galaxies are clearly bluer (and of higher surface brightness) than those of the green and red subsamples.

 All of the analysis carried out up to this point has used the original SDSS imaging of the GAMA sample. Indeed, all of the current GAMA data products derive their optical analysis from these data. More recently the KiDS \citep[Kilo Degree Survey, ][]{DeJong2017} has imaged the GAMA equatorial regions in multiple optical bands to a superior depth and spatial resolution. We have used these data to create pseudo true colour images for the galaxies in our sample, using the $g-$ and $r-$band data for each source. We create cutouts of size 5x5 Kron radii centered on each galaxy in our sample. A synthetic intermediate band is constructed by taking the arithmetic mean of the $g-$ and $r-$band imaging, and the resultant 3 images are arcsinh scaled and coadded following the methodology of \cite{Lupton2004}. Images for a representative sample of our red, green and blue objects with two identified components are shown in Fig. \ref{tci}. These images serve to illustrate the results derived statistically in Figs \ref{bulge_mag},  \ref{bulge_col} and \ref{disk_col}.  Note particularly the close visual correspondence between the bulges in all three subsamples and the way the disks fade from a comparatively high surface brightness blue state to a fainter one with colour similar to that of the bulge between the subsamples. All of this clearly suggests that evolution across the green valley is dominated by a change in the {\it observed} disk properties as they fade and become redder.

We cannot rule out a small fraction (of order $10\%$) of the green population being rejuvenated, previously red galaxies with some recently restarted star formation. Indeed Holwerda (in prep)  found that about 10$\%$ of XUV-selected galaxies (interpreted as galaxies which have recently been ``refuelled" and able to resume star formation) have colours that place them in the green valley. Nevertheless, our results indicate that such objects cannot be the dominant green valley population.

\subsection{Visual morphologies}
\label{LOWZ}
Up until this point, our analysis has focussed on the statistics of the structural  properties of a mass-selected sample of objects at $0.1<z<0.2$. With this, we could demonstrate global differences and similarities between the three colour-based subsets. In order to explore what the detailed morphologies might tell us about the types of objects found within the green valley, we briefly consider a lower redshift ($z<0.06$) sample of GAMA galaxies otherwise selected in the same way as our main sample. Restricting ourselves to these redshifts allows us to incorporate the results of a further pre-existing GAMA-based study into our analysis.

\cite{Moffett2016} visually classified a complete sample of GAMA sources\footnote{A small number of objects did not have types in the original catalogue and were classified on the same scheme for the current work.}  with secure redshifts $z<0.06$ drawn from the same three equatorial fields as used in the present work. We can select from these those systems that meet our colour and mass criteria to explore the morphologies of  systems with the same characteristics as our main sample.  Details of the classification method, involving a consensus   among multiple human classifiers, are given in \cite{Moffett2016}. Results for our colour subsamples are shown in Table \ref{morph_tab}. Unsurprisingly given the results in the earlier sections, the large majority of green galaxies are placed in categories that have substantial bulges. Most of these also have significant disk components and, for this redshift and mass range, there are essentially no bulgeless objects (Sd or Irr). We note that \cite{Bait2017} come to a similar conclusion from a study of de Vaucouleurs T-types as a function of colour in the \cite{Nair2010} sample drawn from SDSS.   The significant difference in the E/S0/Sa fractions between blue and green subsamples again implies that if the green subsample is dominated by objects transitioning from the blue cloud, this transition takes place, and the galaxy turns green, only after a significant bulge component has been formed. The transition may perhaps be associated with disk instabilities  \citep{Porter2014}, as discussed in a similar context to ours by \cite{Brennan2017}.
We note that the dominant morphological classification in our sample of green galaxies is S0/Sa and in some previous work \cite[e.g.][]{Schawinski2014} this intermediate population has been excluded from analysis. 

We can use the same KiDS data set as earlier to produce pseudo true-colour images of this sample (Fig. \ref{tci_low}). Unsurprisingly, this shows the same change in colour and apparent morphology across the blue, green and red subsamples as for our main $0.1<z<0.2$ sample, albeit at a higher physical resolution due to the shorter distances to the objects, again supporting the idea that the apparent morphological change across subsamples is related to evolution in the disk properties. We will expand on the morphological study of this lower redshift sample,  looking  at more detailed substructures such as bars, rings and lenses, in a forthcoming paper.

\begin{figure}
\includegraphics[width=\linewidth]{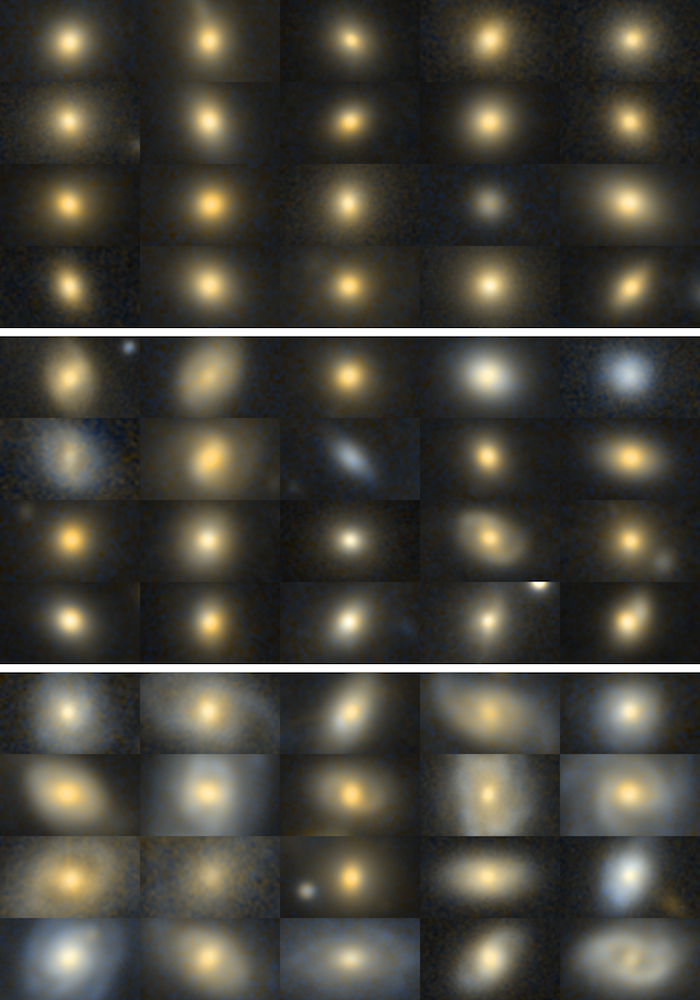}
\caption{ Pseudo true-colour images of  randomly selected examples of our red (top), green (middle) and blue (bottom) subsamples used in subsection \ref{BDD}.  Each image is of 5 x 5 Kron radii and the image is generated using the KiDS $g-$ and $r-$ band images as described in the text with all treated identically in terms of the scaling of their apparent colours. Note that these images are not corrected for each galaxy's intrinsic reddening. The varying appearance across each of the three subsamples reflect the quantitative results derived in Figs. \ref{bulge_mag},  \ref{bulge_col} and \ref{disk_col} and are consistent with the disk component fading and becoming redder across the subsamples.}

\label{tci}
\end{figure}

\begin{table}
\begin{tabular}{lccccc}
Colour & No. & E & S0/Sa &Sab-Scd&Sd-Irr\\
\hline
Red &386 & 30\%&65\%&4\%&$1\%$\\
Green &122&12\%&60\%&27\%&$1\%$\\
Blue &201&3\%&27\%&68\%&2\%\\
\end{tabular}
\caption{ Distribution of morphologies classified as in \citet{Moffett2016}  for $z<0.06$ galaxies in the mass and colour ranges noted in Section \ref{samp}, but with no restriction  (in this one instance) on apparent axial ratio. Applying such a restriction would preferentially remove edge-on disk-dominated systems. }
\label{morph_tab}
\end{table}

\section{Discussion}

\subsection{Disk fading}

\begin{figure}
\includegraphics[width=\linewidth]{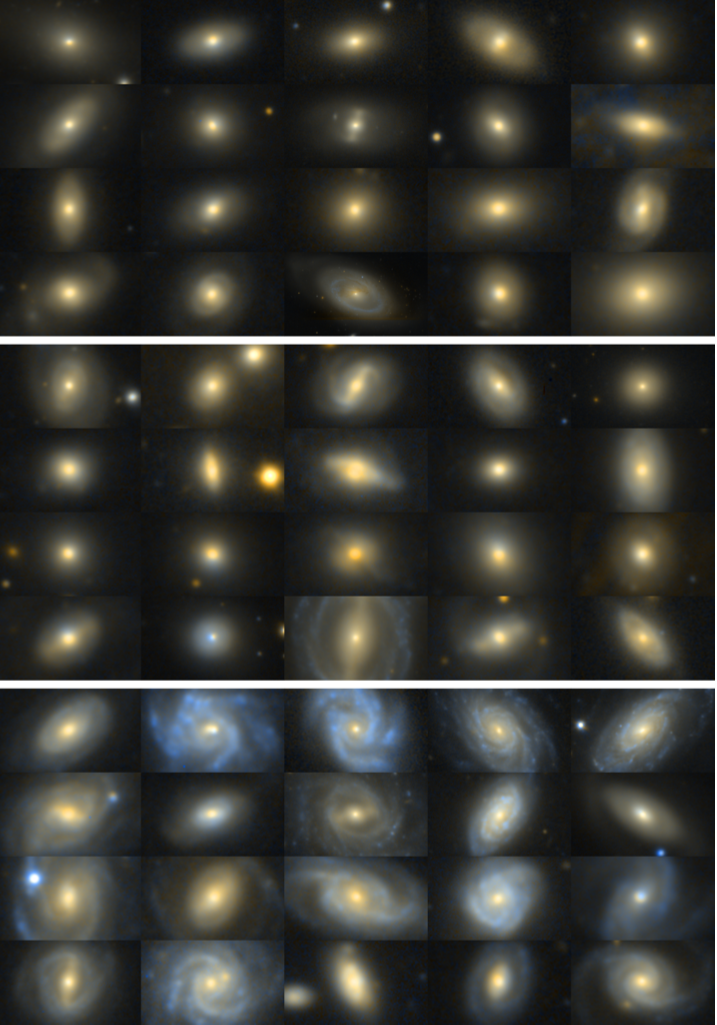}
\caption{ Pseudo true-colour images of  randomly selected examples of our low redshift ($z<0.06$) red (top), green (middle) and blue (bottom) subsamples used in subsection \ref{LOWZ} and drawn from Moffett et al., (2016).  Each image is created identically to those in Fig. \ref{tci}. The behaviour  shown in these subsamples is evidently consistent with that seen in the higher redshift subsamples, but seen at a higher physical resolution (by approximately a factor $\sim 3$ due to the difference in the typical redshifts of each sample). }

\label{tci_low}
\end{figure}

Given the correlation between the intrinsic colours of their stellar populations and their specific star formation rates (e.g. Fig. \ref{SSFR}), there is evidently still ongoing star formation within the green population  albeit at a level an order of magnitude lower than in the blue cloud.  The measured SSFRs imply that a long-term steady state in star formation is incompatible with an intrinsically green stellar spectral energy distribution at the masses considered here unless there is very fine tuning of the fuelling (and therefore star formation) rates over many Gyrs. The bulk of these green galaxies are therefore either those transitioning from the blue cloud  towards the red sequence as star formation declines or are (temporarily) rejuvenated red galaxies. At least one study of the stellar population of the Milky Way disk \citep{Haywood2016} suggests that low-level rejuvenation lasting several Gyr is possible after a temporary, 1-2~Gyr, dip in star formation \citep[though see][]{Snaith2015}. However,  the KiDS imaging of the green objects (Figs. \ref{tci} and  \ref{tci_low}) generally shows little evidence for localised star formation as might be expected in the case of recent rejuvenation for the green galaxies.

An important aspect of the first scenario is whether the colour changes as a largely dead spheroid grows to eventually dominate over an actively star forming disk, or whether the spheroid is already in existence before exit from the blue cloud and the colour change is due to declining star formation within a (fading and reddening) disk. Our single component S\'ersic analysis strongly supports the second of these alternatives \citep[as also found by] []{Carollo2016}. In the $K-$band, where the surface brightness best traces the underlying stellar mass profile, the profile parameters for the green galaxies are typically far closer to those of the red galaxies than those of the blue galaxies (Figs. \ref{nvre} and \ref{hist_sersic}). However, in the optical ($g-$band), where the surface brightness profile has a far higher sensitivity to radial mass-to-light ratio variations due to younger populations \citep[cf.][]{Goddard2017}, the green population appears to be intermediate between the red and blue, reflecting the influence of disk components.

This agrees with and strengthens the conclusion of \cite{Fang2013} that at fixed stellar mass, the comparatively narrow range of (high) central mass densities of green galaxies  is very similar to that found for red systems, whereas blue galaxies have a much wider range for this parameter, extending to significantly lower values.  Together, these results clearly imply that for our studied stellar mass range among current-day galaxies, a significant bulge or spheroid component is in place {\it before} the object enters the green valley. \cite{Vandokkum2013} demonstrated that for galaxies in the mass range considered here, the growth of the central stellar population - which we can take to be the bulge - significantly slows below $z\sim 1$ and subsequent star formation is concentrated in the disk, consistent with our findings.

This still allows for either of the two distinct disk star formation histories noted above, fading or rejuvenation.
On the other hand, the results in Section \ref{morpho} appear to point clearly to the fading scenario being the dominant one. The clear variation in disk colours going from blue to red subsamples  argues for a fading disk. The dominance of the E/S0/Sa morphological classifications for the green subsample in Table \ref{morph_tab} clearly reiterates that a mature strong spheroid component is the norm for the green population, even in the presence of a disk component.  

Looking at individual images of green valley galaxies, disks generally appear devoid of obvious star forming regions or even clear spiral arms in most cases and are usually clearly redder than those of the blue sample, reflecting the balance of their  Hubble types in Table \ref{morph_tab}. They also do not regularly show  small-scale regions or patches of star formation, which might be expected in a rejuvenating system, rather the bulk have visual morphologies consistent with  fading throughout the disk \citep[cf.][]{Belfiore2017}. Very few show any evidence for strong interactions or mergers. On the other hand, bars, rings and lenses are frequently seen and the presence of the such features are often taken to be symptoms of gravitational instability in disks \citep[e.g. ][]{Buta2012}. Simulations \citep[e.g.][]{Athanassoula2013, Kim2016} indicate that such features generally take several Gyr of steady evolution to form.

\subsection{Environment and timescales}

The results in Section \ref{env} on the relative fraction  of red, green and blue galaxies as a function of environment can be interpreted as a reflection of the timescale to transition across the green valley, assuming that the predominant evolutionary path is from blue to red and transition only happens once, rather than some stochastic rejuvenation process.  At the simplest level, the approximately constant green fraction (15-20 per cent) across the whole range of environments probed here can be equated with the fraction of the lifetime of a typical galaxy within the green phase. Estimates for the lifetimes of GAMA galaxies have been determined by \cite{Taylor2011} from SED fitting assuming exponentially declining star formation histories, with the e-folding time and time since the initiation (peak) of star formation (the `age') as free variables. For our subsamples, ages are independent of environmental richness over the range investigated and are typically 7-8~Gyr for old systems.  The fraction of galaxies which are green therefore implies a timescale for crossing the green valley of order  1-2 Gyrs, rather than hundreds of Myr.  On the other hand, if a subset of the red galaxies can be characterised as the result of early-time monolithic collapse or generally formed on a very short timescale and did not spend a significant fraction of their lifetimes as (blue) star forming galaxies, then this implies a longer timescale for traversing the green valley. Only if a minority of the green galaxies are truly transitioning  and the rest are intrinsically green for some other reason (seemingly ruled out above), can the transition timescale be appreciably shorter. Evidently, if there are a range of timescales for traversing the green valley, any characteristic timescale measured from  the relative fractions of galaxies by colour will  reflect the dominant and longest of these \citep[see arguments in][]{Martin2007}. At an extreme, a subpopulation that transitions on much shorter timescales would not contribute to the green fraction. If such a population exists it either must be relatively rare, or must be balanced by (perhaps the same) galaxies crossing the green valley in the other direction to avoid emptying the blue cloud.  If so, we might not expect these galaxies to have the relatively relaxed and smooth morphologies while they have red colours, as seen in Figs. \ref{tci} and \ref{tci_low}

 We note that the latest IllustrisTNG simulations predict a comparable timescale ($\sim 1.6$ Gyr) for galaxies to traverse the green valley \citep{Nelson2017}, with the primary driver of colour transition in this simulation being  low accretion rate black hole feedback. The same simulation indicates a minor r\^ole for rejuvenation in populating the green valley with most simulated green valley occupants transitioning from blue to red populations at essentially constant mass, in agreement with our earlier assumption.

Given that any signature of strong interaction or major merging between galaxies is expected to last several crossing times (a good fraction of a Gyr), the very low incidence of such signatures in the green subsample indicates that such events are not responsible for triggering the bulk of transitions into and across the green valley. \cite{Blanton2006} already came to a similar conclusion by statistical comparison of the blue clouds in the DEEP2 and SDSS  surveys.

The lack of variation in green fraction with environment appears to indicate little environmental effect on  timescale for traversing the green valley, at least over the mass range investigated here. While Fig. \ref{sigma5_hist} shows that green galaxies have local environmental richnesses intermediate between those of the blue and red populations,  the increasing ratio of green to blue galaxies with environmental richness clearly signifies a role for the environment in instigating the transition across the green valley \citep{PengY2010}. Again, we can simply interpret the ratio of green to blue galaxies as a direct measure of the ratio of time spent in each phase as a function of environment. If so,  in relatively low density environments, the blue phase typically lasts twice as long as the green while in the richest environments, the blue phase is shorter (for our mass range). This indicates a r\^ole for the environment in decreasing the ongoing star formation rate soon after a blue galaxy either enters or is formed in a rich environment, changing its colour.  This  assumes that all blue galaxies are equally capable of transitioning through the green valley.  The survival time for blue galaxies in the richer environments probed here (the time before they become green or red) cannot be much shorter than $\sim 1$~Gyr given the blue fraction in even the richest of these environments. Too short a lifetime would require the current infall rate to be far higher than it must have been historically, given the total number of galaxies in each richness bin \cite[see also][]{Wijesinghe2012}.

Regardless of the transition timescale, the radial mass distributions indicated by the $K-$band S\'ersic fits  in Section \ref{phot} and the central surface densities of green galaxies reported by \cite{Fang2013}, imply that most blue galaxies would have to evolve their radial mass distributions before they can decrease their specific star formation rate to a level where they enter the green valley. So, at least for this mass and redshift range, most green valley galaxies appear to have  developed radial stellar mass profiles comparable to red galaxies {\it before} transitioning  across the green valley.

\subsection{Gas supply}
The variation in intrinsic colour and SSFR between galaxies can be explained in terms of a simple picture where  a galaxy starts to decrease its SSFR as  the reservoir of immediately usable (cold) gas runs low. This could be either simply through long-term depletion of  the available  reservoir, with inflow into the reservoir being insufficient to maintain it at previous levels,  or prematurely through an environmental process that removes or suppresses infall of a significant fraction of the cold gas from the larger environment. Complete HI observations of the GAMA survey are not yet available, but we are able to use the results of the  GALEX Arecibo SDSS survey (GASS) \citep{Catinella2010,Catinella2012, Schiminovich2010}  of SDSS galaxies with $\rm{log}(M_*/M_{\odot})>10$  and $0.025<z<0.05$ to infer how gas supply and star formation are related. GASS   demonstrated that regulation of the gas supply to galaxies is central to controlling their star formation \citep[see also, e.g.][]{Peng2014}. Over the mass range studied, the mean value of the ratio of  star formation rate to HI gas mass - the star formation efficiency (SFE) -  appears to vary little with  stellar mass or NUV$-r$ colour (or equivalently SSFR), albeit with a large scatter. The NUV$-r$ colour does correlate with the ratio of HI mass to stellar mass \citep{Catinella2012, DeVis2017}. This latter correlation indicates that typical green valley galaxies have a $M_{HI}/M_*$ ratio at least an order of magnitude lower than typical blue cloud galaxies and it is this comparative lack of gas  in a galaxy's immediate environment that drives the relative star formation rates. 

From a theoretical perspective, a model where the growth of a bulge (e.g. due to disk instabilities) is accompanied by the simultaneous growth of a central black hole could well explain this type of correlation. When itself sufficiently-fuelled, the black hole can generate AGN feedback which is well understood in theoretical models to remove gas from and/or reduce gas inflow rates onto a galaxy thereby reducing its star formation rate \citep[e.g.][]{Booth2010,Gabor2011}.  This has been investigated specifically in the context of green valley galaxies by \cite{Terrazas2017}, though also see \cite{Eales2017b} for possible arguments against this being the dominant method of suppressing star formation at recent epochs.

The mean SFE translates to a timescale of $\sim 3$~Gyr for exhausting the available HI  at the current star formation rate (assuming complete conversion of atomic gas to stars), consistent with typical gas consumption timescales \citep[e.g.][]{Hopkins2008}. This is an order of magnitude more than the typical dynamical time for reasonably massive star forming galaxies, so agrees with the efficiency of HI consumption required by the Kennicutt-Schmidt law \citep{Kennicutt1998}.  Although the available mass of gas for a given galaxy can vary stochastically with time (e.g. through inflows and outflows), statistically it is reasonable to assume that the ratio of gas to stellar mass decreases over time as gas is turned into stars. We know that this is true on a global scale from the strong decline in the cosmic star formation rate and the old stellar population ages of typical massive red sequence galaxies \citep[e.g.][]{Thomas2010}. The typical blue galaxy is then a smaller fraction of the way though its supply than a green or red one. If left alone a galaxy with insufficient  cold gas inflow or cooling from a warm/hot phase would eventually  diminish  its cold gas supply sufficiently to naturally transition from the blue cloud through the green valley to the red sequence. This will evidently be the case if the cosmological inflow  of gas onto the galaxy \citep[e.g.][]{Keres2005} is cut-off, but even if it is not, observations suggest that in practice inflow and cooling rates are insufficient to maintain a `normal' (blue) level of star formation \citep{Hopkins2008,Sancisi2008}. A clear implication of the lack of large-scale variation of  SFE with colour is that  while the stages where the gas content and SFR are low (due to the galaxy having used up most of the initial gas, with little or no current replenishment; \cite{Larson1980} {\it et seq.}) can be prolonged over several Gyrs \citep[see also][]{Eales2017b}, the galaxy colours will asymptote to the red sequence, i.e. the galaxies will `pile-up' at red colours \citep{Eales2017a}. If the gas supply is disrupted or sufficiently diminished, then as long as the SFE is maintained, the SSFR will need to drop rapidly to be consistent with the GASS results.

\subsection{ A simple model}

We can use these simple concepts of  regulating star formation through the gas supply to interpret our results. The isolated field galaxies in Table \ref{tab_env} could simply demonstrate the behaviour of galaxies working their way through their cold gas supply with the green and red subsets being those that are close to, or at, the point where the available cold gas and the rate at which it is replenished only allow star formation at very low rates.

Those galaxies in the same table with richer environments (from pairs upwards) show a decreasing blue fraction. This could be interpreted as the result of their environment acting to prematurely reduce the cold gas supply to an increasing subset, presumably through galaxy-galaxy, or galaxy-intra group medium interactions (strangulation, ram pressure stripping etc. - see \citealt{Grootes2016} for an extensive discussion). The velocity dispersions \citep[from][]{Robotham2011} of the less rich environments vary from typically $\sim 200$ to $\sim 350$ kms$^{-1}$ for systems with multiplicities up to 20 galaxies, but can be $\sim 500$ kms$^{-1}$ or more for the richest subset in Table \ref{tab_env}. Consequently, significant galaxy-galaxy interactions are perfectly viable in all but the richest bin. However, significant interaction with a hot IGM is only likely in a subset of the richest environments. Given that we see a continuous change in the ratio of green to blue galaxies with increasing environmental richness, clearly both  mechanisms are likely to be implicated in disrupting or diminishing the cold gas supply depending on environmental richness. 

Any process that reduces the overall available gas supply does not necessarily affect the instantaneous star formation rate, but acts to shorten the time that the galaxy can maintain the SSFR required to remain in the blue cloud. Star formation does not need to be immediately quenched, the process simply artificially ``ages" the galaxy by bringing closer the time that it starts to run low on gas \citep[cf.][]{vonderLinden2010}.

Given all of the above evidence, we can explain the bulk of the green valley population in this mass and redshift range if it is largely made up of early type disk galaxies  that are close to exhausting their available fuel for disk star formation, either through age or it having been prematurely depleted. At an earlier stage of their evolution, the galaxies are either supplied directly   from infalling/cooling gas at a rate high enough to maintain the ``normal" SSFR for star forming galaxies at that epoch, or had access to a large enough pre-existing reservoir of such gas \citep[as in gas regulation models of galaxy evolution, e.g.][]{Lilly2013}. The reservoir acts as a buffer, allowing this level of star formation even if the replenishment rate from infalling or cooling material was significantly lower. Over time, either the rate of direct supply has dropped or the reservoir has been diminished  and star formation can  then only proceed at a rate dictated by that of  the infalling or cooling material.  Any replenishment of the cold gas reservoir, or direct supply of gas  to a galaxy at the redshifts studied here is at a rate below that required to maintain a blue-cloud level of SSFR \citep[e.g.][]{Hopkins2008}. Then, simply over the next 1-2 Gyr the star formation in the disk decreases further leading to a reddening and fading of the disk. 

Over time the galaxies become apparently more and more spheroid dominated in terms of their optical appearance, though in reality their radial stellar mass distributions do not have to change very much (they already contained a significant bulge, by mass). The apparent evolution is due to an increase in mass-to-light ratio in the disk due to its fading. The extremes of the $M_*/L$ distribution in Fig. \ref{moverl} represent those of pure disks and pure bulges, so imply that a disk can fade by 1.5 magnitudes or more in optical flux and surface brightness and even by 0.5 magnitudes in the $K-$band relative to any bulge component. Clearly, as this happens a galaxy will appear increasingly spheroid dominated and in the optical its morphological classification is likely to change  to an S0 or even an E as the apparent (i.e. optical luminosity) bulge-to-disk ratio increases by a factor $\sim 4$. Given that the galaxies will have an intrinsic range of bulge to disk mass ratios we should not necessarily expect to see a clear change in the statistics of the S\'ersic fits to the optical surface brightness distributions with intrisic colour within the green valley.  Any particular galaxy may have a higher bulge mass fraction and a disk that has faded less than another of the same mass and therefore appear redder even though it is less far through the disk fading process. Nevertheless, we do see a change in average $g-$band S\'ersic index $\Delta n_g \simeq 1$ between the bluest and reddest thirds of our green subsample, consistent with the typical galaxy's disk fading as it traverses the green valley. Consequently the typical bulge to disk stellar mass ratio of the green galaxies cannot vary arbitrarily. This in part could be a selection effect as the extremes of the distribution would fall into the blue cloud or red sequence.

We therefore expect for green valley galaxies with the same bulge to disk stellar mass ratios, the reddest objects to be both the furthest along in this process with consequently the lowest gas to stellar mass ratios and with higher apparent optical bulge to disk (light) ratios than their slightly bluer counterparts.  \cite{Schawinski2014} demonstrated that those green valley galaxies classified in Galaxy Zoo as being early type (optically very bulge dominated) had redder $NUV-u$ colours than those classified as late type (clearly disk dominated). Given that the $NUV$ luminosity reflects the star formation history in the past 10-100 Myr, this implies less recent star formation  in the early types even if the intrinsic $u^*-r^*$ colours (sensitive to star formation on Gyr scales) are the same as for the green valley late types.  This was shown to be consistent with different timescales for reaching the red sequence  for two types of galaxies, with the early types evolving in colour much faster than the late types, as expected if their gas supply was catastrophically disrupted.  

Here we note that such behaviour is also consistent with the  early types being associated with the late stage of continuous disk fading which we see here, even if the bulge to disk stellar mass ratio does not change as a galaxy traverses the green valley.  Our analysis in  this regard benefits from including all morphological types as measured by their S\'ersic indices and so is able to trace a continuum of evolution in colour and morphology.
If we use the same morphological classification as that used in \cite{Schawinski2014}, Table \ref{morph_tab} indicates that only 15 per cent of our green sample would be considered early type and 25 per cent late type (if we assume the S0/Sa class corresponds to the indeterminate-types in the Galaxy Zoo sample). The full Galaxy Zoo sample itself  contains 45 per cent intermediate types, though it has a lower mass limit (by an order of magnitude) than our sample. So, the significant majority (85 per cent) of the green valley population explored in this work have identifiable disks and at least a fraction of the earliest types could be the ``end states" of galaxies with disks that have faded sufficiently that optical classification ignores them \citep[see also ][]{Carollo2016}. So while it is entirely possible that there is a population of early types in the green valley that follow a different evolutionary pathway across it to that of the objects with more prominent disks, a subset of the early types could simply be the end state of the disk fading scenario discussed above, particularly for objects with higher than average bulge to disk stellar mass ratios.

\section{Conclusions}

We selected galaxies from the GAMA II survey in the stellar mass range $10.25<\rm{log}(M_*/M_\odot)<10.75$   at $z<0.2$ in order to study how current-day galaxies transition between the blue cloud and red sequence. The mass range was chosen to ensure that significant numbers of galaxies over the full range of stellar population colours were present in our sample. Objects were split into blue, green and red subsamples based upon the intrinsic (dust-corrected) rest-frame $u^*-r^*$ colours of their stellar populations, with the green objects defined to fall between the blue cloud and red sequence in a colour-stellar mass diagram.

As expected, we find that the green galaxies  have  $g-$band surface brightness profiles as parameterised by S\'ersic index and effective radius  midway between those of blue cloud and red sequence galaxies of the same stellar mass. However, their $K-$band surface brightness profiles (a proxy for the stellar mass surface density)  are far closer to those of red sequence galaxies, even in the case of the bluest green valley galaxies. The implication is that the buildup of the stellar mass profile largely precedes entry into the so-called green valley, rather than occurring as the galaxies traverse it.  Bulge-disk decomposition of a subset of our sample indicates that the bulges of red, green and a subset of the blue galaxies are similar in $K-$band luminosity. Over this mass and redshift range, most  green galaxies already have their bulges in place and we see no clear evidence for significant bulge growth within the green valley. The fraction of blue galaxies with similar structures are plausibly the green galaxies' immediately prior stage, with the bulges formed but star formation not yet significantly reduced.

Across all environments studied here (from isolated galaxies to moderately rich groups), the fraction of galaxies with a given environmental richness that are green remains approximately constant at 15-20 percent.  Assuming at best a minority of these  galaxies are (temporarily) re-invigorated star forming previously red galaxies, this fraction indicates a timescale for traversing the green valley from blue to red of order $1-2$~Gyr, in agreement with previous studies which derived the timescale through different means. 

Green galaxies are found in environments intermediate in {\it local} richness (measured by surface density or number of near neighbours) relative to typical blue and red population members, which as expected occupy lower and higher density environments, respectively. Notwithstanding the similar overall green fractions throughout, the  ratio of green to blue galaxies increases significantly with {\it larger scale} environmental richness (from $<0.5$ in the field to $>1$ in moderate groups) implying that while environment does not particularly affect the timescale for crossing the green valley, it can trigger or enable the process by which the blue galaxies start to transform. It implies a survival time for the blue galaxies upon entering richer environments that is shorter than the typical green valley lifetime. 

The lack of obvious signatures of  mergers and  interactions in green galaxy  morphologies  indicates these play little or no part in triggering most transformations. This strengthens  the conclusion in \cite{Blanton2006} and more recently \cite{Vandokkum2013} that blue galaxies simply evolve  with a decreasing star formation rate between $z=1$ and today. 

The colour change from blue to red in objects where two components (bulge and disk) could be separated is driven mainly by colour change in the disk. 

Morphologically, green galaxies are predominantly classified as S0/Sa types, with a substantial minority classed as later spiral types, in contrast to blue galaxies which are largely late types. The large majority of the green valley objects show both bulge and disk components. Pure spheroids are a small minority. While this subpopulation may well contain systems that evolve quickly through the green valley as demonstrated by \cite{Schawinski2014}, many may simply be the end-point of the disk fading process in systems that would have been classified earlier in their evolution as early type spirals with relatively high bulge fractions.

Images of  green galaxies do not usually show clear spiral arms and we rarely see features with colours that indicate obvious ongoing star formation. The dominant population is consistent with the disk fading hypothesis rather than rejuvenated star formation in either disk or bulge. Substructures such as rings appear to be quite frequent and we  consider these in detail in a separate publication.

 As expected, the rest-frame $u^*-r^*$ colours of the stellar populations correlate with specific star formation rate and this in turn is known to correlate with HI gas to stellar mass ratio \citep{Schiminovich2010}. Consequently, our results can be interpreted as the bulk of our green sample being galaxies that  have sufficiently diminished their original cold gas supply which is now being replenished at a rate lower than that required to maintain previous levels of star formation.  As the same prior work also demonstrates that the star formation efficiency (star formation rate per unit HI mass) does not vary systematically across galaxies of different colours, our timescale for traversing the green valley is simply a consequence of this. 

 \section*{ACKNOWLEDGEMENTS}

GAMA is a joint European-Australasian project based around a spectroscopic campaign using the AngloÐAustralian Telescope. The GAMA input catalogue is based on data taken from the Sloan Digital Sky Survey and the UKIRT Infrared Deep Sky Survey. Complementary imaging of the GAMA regions is being obtained by
 a number of independent survey programs including {\it GALEX} MIS, VST KiDS, VISTA VIKING, {\it WISE}, {\it Herschel}-ATLAS, GMRT and ASKAP providing UV to radio coverage. The VISTA VIKING data used in this paper are based on observations made with ESO Telescopes at the La Silla Paranal Observatory under programme ID 179.A-2004. GAMA is funded by the STFC (UK), the ARC (Australia), the AAO, and the participating institutions. The GAMA website is http://www.gama-survey.org/. This work made extensive use of TOPCAT \cite{Taylor2005} and STILTS \cite{Taylor2006} software packages, which are supported by an STFC grant to the University of Bristol.

\label{lastpage}

\end{document}